\documentclass[fleqn,useAMS,usenatbib]{mn2e}
\pdfoutput=1
\usepackage{mathptmx}

\usepackage[totalwidth=480pt, totalheight=680pt, a4paper]{geometry}

\usepackage{amssymb,amsmath}
\usepackage{fixltx2e}
\usepackage{graphicx}

\renewcommand{\vec}[1]{\mathbfit{#1}}

\title[Simulations of the non-linear thin shell instability]{Simulations of the non-linear thin shell instability}
\author[A. D. McLeod and A. P. Whitworth]{A. D. McLeod$^{1,2}$\thanks{E-mail: amcleod@astro.ex.ac.uk} and
A. P. Whitworth$^{2}$\\
$^{1}$Astrophysics Group, University of Exeter, Stocker Road, Exeter, EX4 4QL, United Kingdom\\
$^{2}$Cardiff School of Physics and Astronomy, Cardiff University, Queens Buildings,
The Parade, Cardiff, CF24 3AA, United Kingdom}

\begin{document}

\date{Accepted 2013 January 31. Received 2013 January 24; in original form 2012 October 22}

\pagerange{\pageref{firstpage}--\pageref{lastpage}} \pubyear{2013}

\maketitle

\label{firstpage}

\begin{abstract}
We use three-dimensional smoothed particle hydrodynamics simulations to study the non-linear thin shell instability (NTSI) in supersonic colliding flows. We show that for flows with monochromatic perturbations and for flows with white-noise perturbations, growth speeds approximate quite well to the analytic predictions of \citet{Vishniac1994}.

For flows with subsonic turbulence, growth speeds match \citeauthor{Vishniac1994}'s predictions only at short wavelengths where the turbulence is weaker. We find that supersonic turbulence, of a lower Mach number than the colliding flows, completely suppresses the NTSI. Our results provide a diagnostic for identifying the presence of the NTSI in colliding flows with turbulence.
\end{abstract}


\section{Introduction}

Colliding flows of gas are common in the interstellar medium and lead to the formation of dense, shock-compressed layers \citep[e.g. ][]{DraineMcKee1993}. Since stars form from dense gas, colliding flows may be important triggers of star formation. Colliding flows can occur at the confluence of expanding H\,\textsc{ii} regions \citep{ElmegreenLada1977}, stellar-wind bubbles \citep*{StevensBlondinPollock1992} and supernova remnants \citep{Williams_et_al1997}, or from large-scale galactic flows (\citealp{Bonnell_et_al2006}; \citealp{VazquezSemadeni_et_al2006}; \citealp*{HeitschHartmannBurkert2008}; \citealp{Hennebelle_et_al2008}). Turbulence, which is believed to exist at a range of scales in the Galaxy \citep[e.g.][]{MacLowKlessen2004}, also creates colliding flows.

In this paper we use three-dimensional smoothed particle hydrodynamics (SPH) simulations to explore the effect of the \emph{non-linear thin shell instability} \citep[NTSI;][]{Vishniac1994} on the development of a shock-compressed layer between two supersonically colliding flows. We seed the inflowing gas with a variety of initial perturbations, and compare the resulting NTSI growth speeds with the analytic predictions of \citet{Vishniac1994}. Previous studies have been limited to two-dimensional simulations \citep{BlondinMarks1996, KleinWoods1998, Hueckstaedt2003}, and only \citeauthor{BlondinMarks1996} attempted to measure the dependence of growth speeds on wavenumber.

\subsection{NTSI}

\begin{figure}
   \centering
   \includegraphics[width=\linewidth]{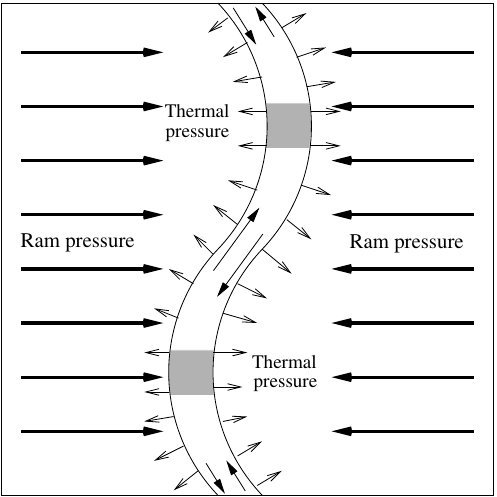}
   \caption[Illustration of NTSI]{The non-linear thin shell instability. Gas flows in from the left and right, confining the layer with ram pressure (large arrows). The dense gas in the layer resists the compression with thermal pressure (small arrows), but this is always normal to the surface of the layer. The misalignment of ram and thermal pressure causes shear in the layer (shown with arrows within the layer), leading to density enhancements (shown in dark grey).}
   \label{fig:NTSI_diagram}
\end{figure}

The NTSI occurs in a dense layer that is already corrugated, as illustrated in Fig.~\ref{fig:NTSI_diagram}. The thermal pressure exerted by the material in the layer always acts perpendicular to the surface of the layer, whereas the ram pressure of the inflowing gas acts along the collision axis. The resulting shear drives material towards the extremities of the corrugation, and this causes the perturbation to grow.

The NTSI was first described by \citet{Vishniac1994}, who considered colliding flows in an isothermal gas. He derived an expression for the growth time $\tau$ as a function of the wavelength $\lambda$, the amplitude $\eta$ and the sound speed $c_\mathrm{s}$,
\begin{equation}
   \tau \left( \lambda, \eta \right) \sim \frac{\lambda^{3/2}}{\eta^{1/2} c_\mathrm{s}} \;.
\end{equation}
It follows that short-wavelength modes grow fastest, and the growth speed is given by
\begin{equation}
   \dot{\eta} \left( \lambda, \eta \right) \sim \frac{\eta}{\tau \left( \lambda, \eta \right)} \sim \left( \frac{\eta}{\lambda} \right)^{3/2} c_\mathrm{s} \;.
\end{equation}

The NTSI is a non-linear instability and can only grow if it is seeded by a finite perturbation with an initial amplitude $\eta_0$ comparable to or larger than the thickness of the layer,
\begin{equation}
   \xi \left( t \right) \approx \frac{2 c_\mathrm{s}^2 t}{v_0} \;,
\end{equation}
where $v_0$ is the speed of the inflowing gas.

\subsubsection{Range of unstable wavelengths}
\label{sec:unstable_wavelengths}

\citet{Vishniac1994} predicts that the NTSI can grow for only a limited range of wavelengths. For a stationary slab in a highly supersonic collision, the large wavelength limit is given by
\begin{equation}
   \label{eq:NTSI_minimum_wavelength}
   \lambda \ll c_\mathrm{s} t \;,
\end{equation}
which states that wavelengths are not excited on larger scales than those coordinated by sound waves. The small wavelength limit is given by the thickness of the layer; a thick layer cannot be effectively corrugated by a small wavelength perturbation, and so
\begin{equation}
   \label{eq:NTSI_maximum_wavelength}
   \lambda \gtrsim 2 \xi \;.
\end{equation}

\subsubsection{Growth to saturation}
\label{sec:growth_to_saturation}

Once the NTSI has been excited at a wavelength $\lambda$, it grows to saturation at a growth speed \citep{Vishniac1994}
\begin{equation}
   \label{eq:NTSI_saturation_speeds}
   c_\mathrm{s} \left( \frac{\xi}{\lambda} \right)^{3/2} \; \lesssim \; \dot{\eta} \; \lesssim \; \frac{\eta}{\lambda} c_\mathrm{s}\;.
\end{equation}
The lower limit is the growth speed of a perturbation with an amplitude equal to the layer thickness, which is the smallest amplitude perturbation unstable to the NTSI. The higher limit can be rearranged to give a growth time limit,
\begin{equation}
   \tau \gtrsim \frac{\lambda}{c_\mathrm{s}} \;,
\end{equation}
which states that the NTSI does not grow in less than the sound crossing time and is another expression of the large wavelength limit of equation~\ref{eq:NTSI_minimum_wavelength}. Equation~\ref{eq:NTSI_saturation_speeds} also implies that the growth speed is not strongly supersonic; perturbations where $\eta \gg \lambda$ will saturate due to the large bending angle.

Initially, smaller wavelength perturbations grow fastest. As the layer thickness grows, smaller wavelength perturbation stop growing and are overtaken by larger perturbations. The fastest-growing mode has a wavelength of approximately the layer thickness, and will subsequently saturate due to the increasing thickness of the layer.

\subsection{Simulation parameters}

We collide flows of isothermal molecular gas of temperature \mbox{10\,K} and sound speed \mbox{$0.19\,\text{km}\,\text{s}^{-1}$}. The gas has an initially uniform density of \mbox{$10^{-21}\,\text{g}\,\text{cm}^{-3}$}. The flows collide along the $x$-axis with a collision velocity $v_0$ of \mbox{$\pm\,3.8\,\text{km}\,\text{s}^{-1}$}, and so Mach number $\mathcal{M}$ of 20. All simulations are three-dimensional, and are periodic in the $y$ and $z$ directions.

The collision box has dimensions $L_x$, $L_y$ and $L_z$ along the $x$-, $y$- and $z$-axes, respectively. We conduct two types of simulation: simulations with a narrow rectangular cross-section and one-dimensional perturbations, and simulations with a square cross-section and either two-dimensional perturbations or turbulence. The narrow cross-section simulations are of a higher resolution than the square cross-section simulations. Table~\ref{table:simulation_parameters} lists the box size for these simulations, together with the simulation end time $t_\mathrm{end}$, mass per SPH particle $m_\mathrm{SPH}$ and total number of SPH particles $N_\mathrm{SPH}$.

\begin{table}
  \centering
  \caption{Simulation parameters}
  \label{table:simulation_parameters}
  \begin{tabular}{@{}lcccccc@{}}
  \hline
  Cross-section&$L_x$&$L_y$&$L_z$&$t_\mathrm{end}$&$m_\mathrm{SPH}$&$N_\mathrm{SPH}$\\
  {}&(pc)&(pc)&(pc)&(Myr)&($\mathrm{M}_{\sun}$)&{}\\
  \hline
  Narrow & 1 & 0.5 & 0.01 & 0.1 & $9.23\times10^{-9}$ & $8\times10^6$\\
  Square & 2 & 0.5 & 0.5 & 0.15 & $1.48\times10^{-6}$ & $5\times10^6$\\
  \hline
\end{tabular}
\end{table}

We impose one-dimensional perturbations which vary across the $y$-axis, and two-dimensional perturbations which vary across the $y$- and $z$-axes. In this work we use dimensionless wavenumbers measured in terms of the relevant box size, which is always \mbox{0.5\,pc}. Wavenumbers are therefore given by
\begin{equation}
   k = \frac{0.5 \, \text{pc}}{\lambda} \;.
\end{equation}
These dimensionless wavenumbers can be converted to physical wavenumbers by multiplying by \mbox{$2 \, \text{pc}^{-1}$}.

We impose three types of initial velocity perturbation on these flows: monochromatic sinusoidal perturbations, white-noise perturbations, and an initial turbulent velocity field. The white-noise perturbations are formed from a superposition of sinusoidal perturbations of equal amplitude and random phase. Monochromatic and white-noise perturbations are only imposed on a small region close to the contact discontinuity between the colliding flows, reducing to zero at a distance $R_\mathrm{pert}$ away from the contact discontinuity by a smoothing function
\begin{equation}
   \label{eq:f_factor}
   f(x) =
   \begin{cases}
      \frac{1}{2} \left[
         \cos \left(\dfrac{x \upi}{R_\mathrm{pert}} \right) + 1
      \right]  \qquad &\text{if } |x| < R_\mathrm{pert} \\
      0        \qquad &\text{if } |x| \ge R_\mathrm{pert}
   \end{cases} \;.
\end{equation}
The region containing the perturbation, and its corresponding momentum, is rapidly accreted onto the layer, after which only unperturbed gas is accreted onto the layer. The instability is then free to grow and decay without interference from other perturbations.

For turbulent perturbations the initial turbulent velocity field covers the entire simulation. The layer experiences a relatively constant input of perturbations as the turbulent medium is accreted onto the layer. Any growth of the NTSI must compete with these incoming perturbations.

\subsubsection{Monochromatic perturbations}
\label{sec:monochromatic_perturbations}

For a single monochromatic sinusoidal perturbation of integer wavenumber $k$ and amplitude $A$, the additional velocity due to the perturbation is given by
\begin{equation}
   \label{eq:IC_1D_vel_pert}
   \delta v_x(x,y,z) = v_0 \, A \; \sin \left( 2 \upi \, k \frac{y}{L_y} \right) f(x) \;,
\end{equation}
for a one-dimensional perturbation. In two dimensions we use a sinusoidal square-grid perturbation given by
\begin{equation}
   \label{eq:IC_2D_vel_pert}
   \delta v_x(x,y,z) = v_0 \, A \;
      \sin \left( 2 \upi \, k_y \frac{y}{L_y} \right)
      \sin \left( 2 \upi \, k_z \frac{z}{L_z} \right)
   f(x) \;,
\end{equation}
where we have integer wavenumbers $k_y$ and $k_z$; for monochromatic perturbations we use $k_y = k_z$. This square-grid perturbation corresponds to a superposition of the four plane-wave modes $\vec{k} = (\pm k_y, \pm k_z)$ with equal amplitude (but varying sign); we consider only the $\vec{k} = (k_y, k_z)$ mode in our analysis. As described in Section~\ref{sec:fourier_transform}, we have
\begin{equation}
   k = \sqrt{k_y^2 + k_z^2} \;,
\end{equation}
and so when $k_y = k_z$ the wavenumber $k_y$ is related to the wavenumber $k$ by $k = \sqrt{2} \, k_y$.

This means that it is not possible to excite the same wavenumbers of one-dimensional and two-dimensional perturbations, as the periodic boundary conditions constrain us to use integer wavenumbers of $k$ for one-dimensional perturbations, and integer wavenumbers of $k_y$ and $k_z$ for two-dimensional perturbations. For the same set of integers, the wavenumbers for the simulations with two-dimensional perturbations are larger, and the wavelengths correspondingly smaller.

\begin{figure}
   \centering
   \includegraphics[width=\linewidth]{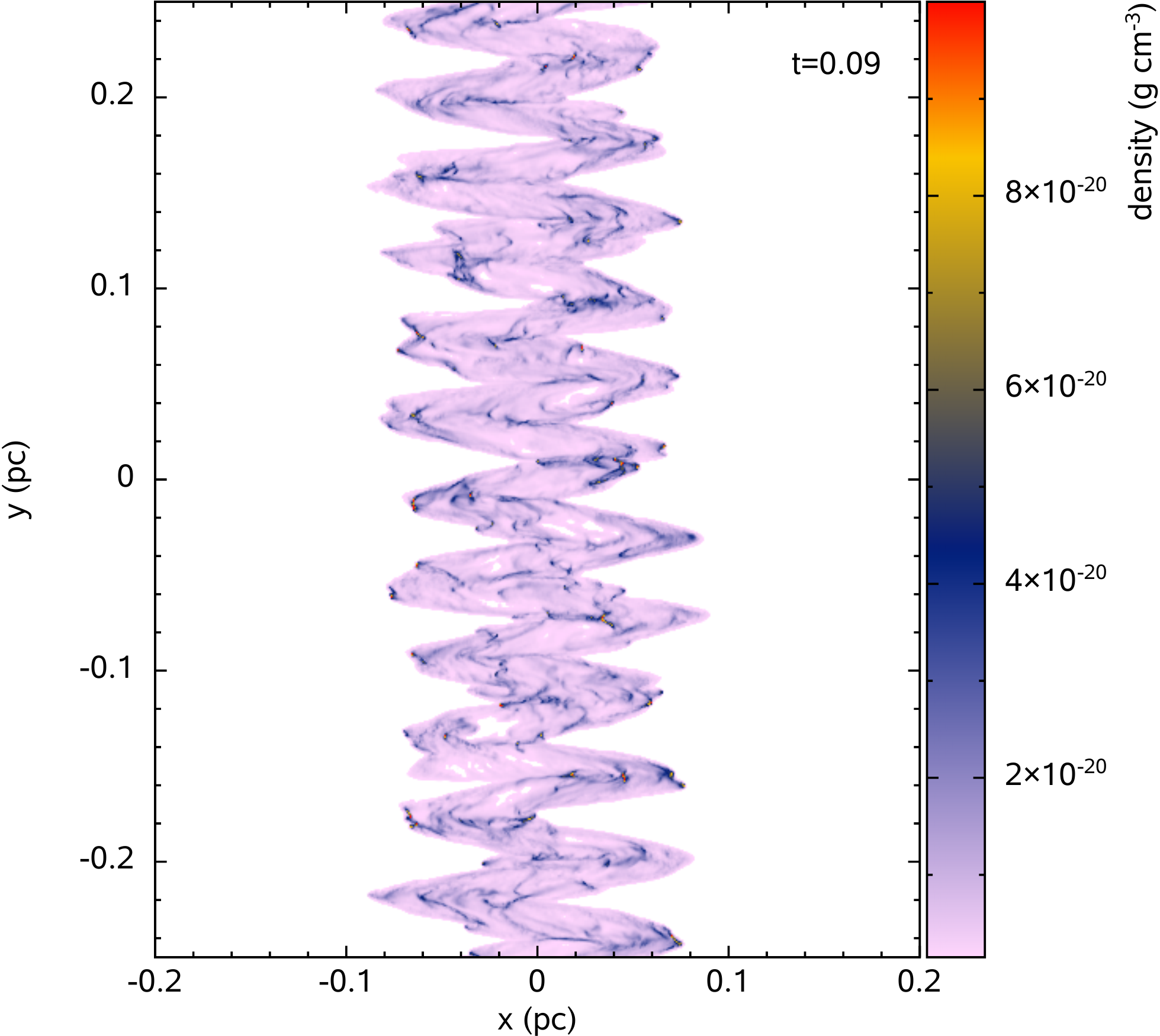}
   \caption{Density cross-section plot of simulation with one-dimensional monochromatic k=12 perturbation after \mbox{0.09\,Myr}. Gas is inflowing from the left and right, driving the NTSI in the dense shock-compressed layer.}
   \label{fig:1D_k12r1_00018}
\end{figure}

We use an amplitude $A$ of 0.2 (except for simulations with two-dimensional monochromatic perturbations where we use an amplitude of 0.5), and a perturbation extent $R_\mathrm{pert}$ (see equation~\ref{eq:f_factor}) of \mbox{0.05\,pc}. We perform simulations using monochromatic wavenumbers $k$ or $k_y$ of 2, 4, 6, 8, 10, 12, 14 and 16 for one-dimensional and two-dimensional monochromatic perturbations. Figure~\ref{fig:1D_k12r1_00018} shows an example of a simulation with a one-dimensional monochromatic perturbation.

\subsubsection{White-noise perturbations}
\label{sec:white_noise}

\begin{figure}
   \centering
   \includegraphics[width=\linewidth]{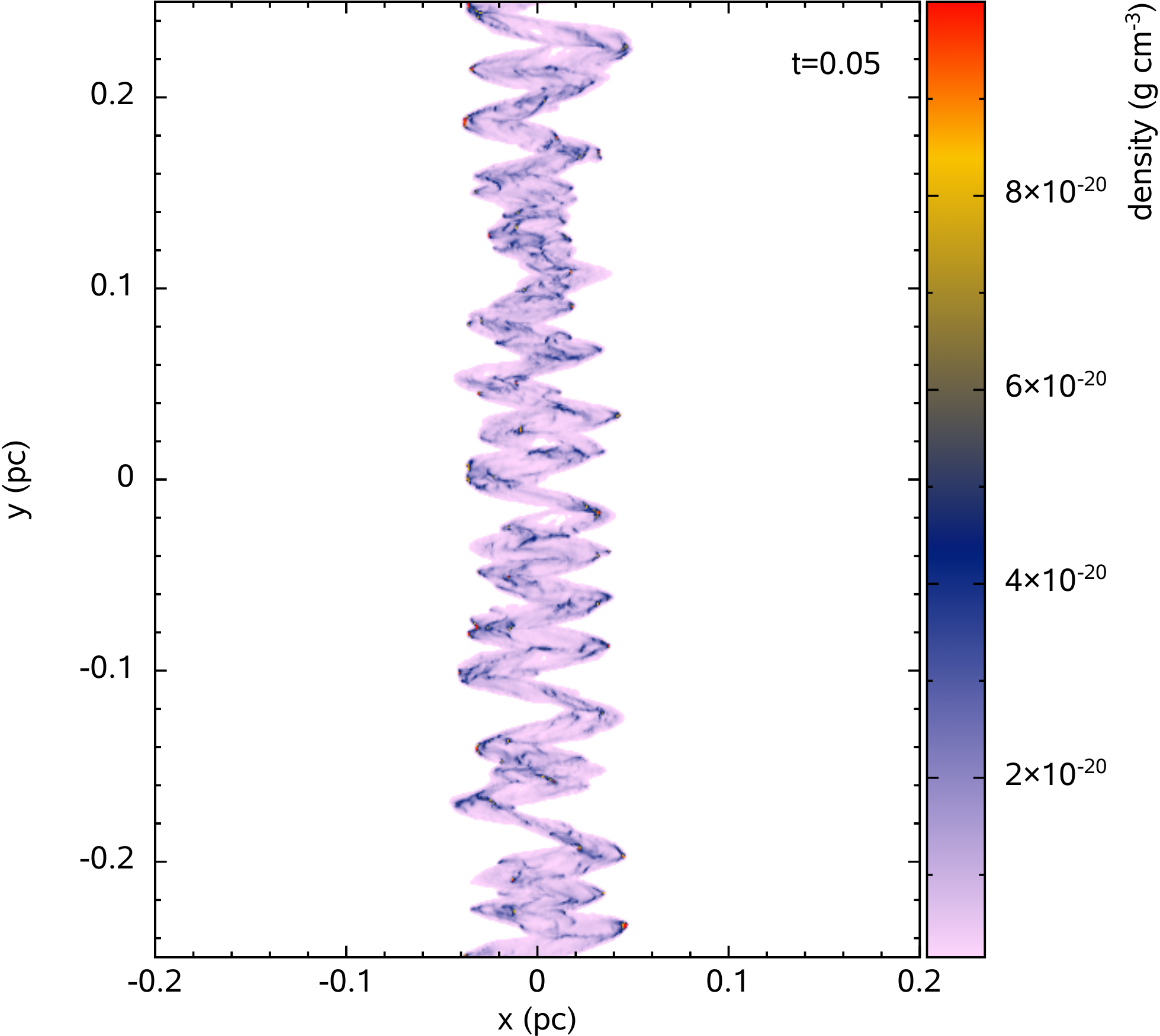}
   \caption{Density cross-section plot of simulation with one-dimensional white-noise perturbations after \mbox{0.05\,Myr}. Gas is inflowing from the left and right.}
   \label{fig:1D_mix1r1_00010}
\end{figure}

White-noise perturbations are generated by the superposition of sinusoidal perturbations between $k=1$ and $k=20$ for simulations with one-dimensional perturbations, and all combinations of $k_y$ and $k_z$ between 1 and 20 for simulations with two-dimensional perturbations. Each wavenumber is given the same amplitude $A$, and a random phase of between 0 and $2 \upi$ chosen from a uniform distribution. We repeat the process to generate ten random sets of perturbations. Figure~\ref{fig:1D_mix1r1_00010} shows an example of a simulation with one-dimensional white-noise perturbations.

\subsubsection{Turbulent perturbations}
\label{sec:turbulence}

In our work we use only divergence-free turbulence generated from a Fourier spectrum with a power-law form $A(k) \propto k^{-2}$ (or equivalently a power spectrum $P(k) \propto k^{-4}$). We generate the cubic and periodic velocity field in a similar way to that described by \citet{MacLow_et_al1998}. The turbulent velocity field is imposed at the start of the simulation and subsequently decays. Turbulent fields are only applied to the lower-resolution simulations with square cross-sections, where they are scaled to fit across the $y$ and $z$ dimensions, and are then repeated across the $x$ dimension as required. This preserves the periodic nature of the $y$- and $z$-axes.

Ten velocity-field realizations are generated. We scale each velocity field to two different average velocities to give a subsonic velocity field with Mach number 0.5 and a supersonic velocity field of Mach number 4. This gives ten subsonic and ten supersonic velocity fields. In both cases the turbulence is only a small perturbation on the colliding flows, which have Mach number 20.

\begin{figure}
   \centering
   \includegraphics[width=\linewidth]{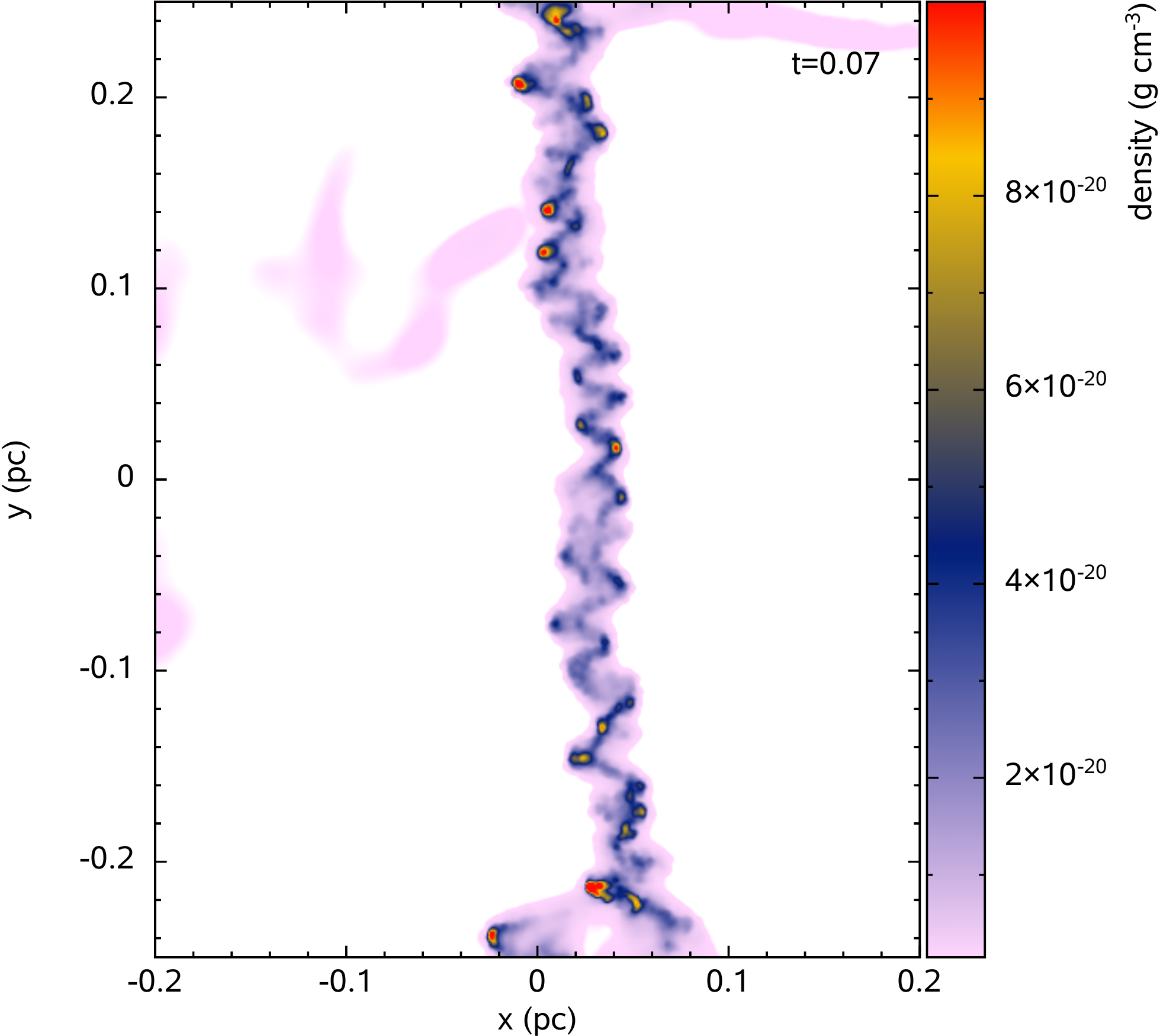}
   \caption{Density cross-section plot of simulation with supersonic turbulence after \mbox{0.07\,Myr}. Gas is inflowing from the left and right.}
   \label{fig:2D_turb1r1_00014}
\end{figure}

Figure~\ref{fig:2D_turb1r1_00014} shows an example of a simulation with supersonic turbulence. The layer appears uniformly displaced to $x>0$ in this figure; this is due to large-scale perturbations in the turbulence. Slices at different $z$ coordinates show displacement or bending of a similar magnitude but both left and right of $x=0$.

\section{Method}

We use the \textsc{seren} \citep{Hubber_et_al2011} code, which solves the equations of fluid dynamics using the SPH method \citep{GingoldMonaghan1977,Lucy1977}. This code includes a conservative `grad-h' SPH scheme \citep{SpringelHernquist2002,PriceMonaghan2004}. We do not include self-gravity or magnetic fields.

\subsection{Creation of initial conditions}

For both the narrow and square cross-section simulations, we generate five realizations of particle positions. We randomly place particles in a periodic box which covers the desired sizes $L_y$ and $L_z$ in the $y$- and $z$-axes, respectively, to ensure there are no regularly repeating patterns of SPH particles along these axes. However, to save computational time, the periodic box has only a fraction of the desired size in the $x$-axis. The particles are allowed to settle under the influence of hydrodynamics and artificial viscosity. Following the settling procedure, the particle positions in the box are repeated along the $x$-axis to generate the desired box length $L_x$.

For the simulations with one-dimensional monochromatic perturbations, each of the monochromatic wavenumbers 2, 4, 6, 8, 10, 12, 14 and 16 is combined with each of the five settled distributions to create a total of forty sets of initial conditions. This is repeated for the simulations with two-dimensional monochromatic perturbations. For the white-noise simulations, there are ten random sets of velocity perturbations; each of the five settled particle distributions are therefore used twice. The same applies for the ten subsonic and ten supersonic simulations with turbulence.

\subsection{Gridding of particle data}

The SPH data is interpolated on to a three-dimensional cubic grid of density using the \textsc{splash} package \citep{Price2007}. This grid is of size $256^3$, is centred at the centre of the collision box, and has sides of length \mbox{0.5\,pc}. Some gas is therefore not included in the grid, and in the simulations with narrow cross-sections the grid covers some empty space across the $z$-axis.

For narrow cross-section simulations, which contain only one-dimensional perturbations, we calculate the average gas density in the simulation box across the $z$-axis. This reduces the three-dimensional grid to a two-dimensional grid of size $256^2$.

Each simulation produces a number of snapshot outputs at regular intervals of \mbox{0.005\,Myr}. The gridding process is repeated for each output snapshot to build a time-sequence of density grids for each simulation.

\subsection{Identification of layer}

Once the data has been gridded, cells containing the dense shock-compressed layer are identified. A density threshold of \mbox{$5\times10^{-21}\,\text{g}\,\text{cm}^{-3}$} is used to identify potential `layer' cells. In simulations containing supersonic turbulence, dense regions can form which are distant from the layer; to remove these from our list of layer cells, we first have to estimate some layer properties.

We calculate the median $x$ position of layer cells along each column of grid cells running along the collision axis. This gives an estimate for the current position of the layer. For each column, we calculate the median absolute deviation of layer cells $x$ position from this estimated layer position; this gives an estimated layer half-thickness for the column. Finally, we take the median of the estimated layer half-thickness for each column to produce an estimated half-thickness for the entire layer.

We then identify discrete groups of connected potential layer cells. We find the average position of each group $(G_x,G_y,G_z)$. We then find the estimated $x$ position of the layer at $y=G_y$ and for square cross-section simulations $z=G_z$. The difference between this position and $G_x$ gives an estimate of the perpendicular distance of the group from the layer $G_\perp$, which we compare with the size of the group along the $x$-axis, $G_\mathrm{L}$.

If $G_\perp$ is larger than $G_\mathrm{L}$ by more than the estimated layer half-thickness, the group is rejected. This is approximately the same as saying that the group is rejected if it does not overlap the layer. All cells within the rejected group are removed from our list of potential layer cells. We find this simple filtering routine, which accounts for variations in the layer position but assumes a constant layer thickness, works well to remove unrelated dense regions formed by turbulence without affecting the identification of the layer.

\subsubsection{Layer position}

Having filtered the layer, we now calculate more accurately the layer position, including only `layer' cells in our analysis. The layer position for each grid column along the $x$-axis is given by the centre-of-mass along that column. This produces a one-dimensional grid of layer positions for narrow cross-section simulations and a two-dimensional grid of layer positions for square cross-section simulations. We use these layer positions as our tracer of bending-mode instabilities present in the layer.

\subsubsection{Layer thickness}

In order to calculate the maximum resolvable wavenumber as a function of time, we need to calculate or estimate the thickness of the layer. We define the layer thickness for each column in $x$ as the total extent of layer cells along $x$ for that column. This includes any empty voids in the layer, as we find this is a better representation of the true thickness of the layer. This is later adjusted to account for the bending angle of the layer.

\subsubsection{Fourier transform}
\label{sec:fourier_transform}

To obtain the Fourier spectrum of perturbations in the layer, we take the Fourier transform of the grid of layer positions, yielding a one-dimensional Fourier spectrum for the narrow cross-section simulations and a two-dimensional grid of Fourier amplitudes for the square cross-section simulations. This is normalized such that values represent the amplitude of perturbations in parsecs.

For the square cross-section simulations, we perform a simple circular averaging technique to convert the two-dimensional grid of Fourier amplitudes, defined at integer wavenumbers $k_y$ and $k_z$, into a one-dimensional Fourier spectrum. This is defined over a similar range of wavenumbers to the Fourier spectrum for the narrow cross-section simulations, but is defined at a large number of fractional wavenumbers rather than only at integer wavenumbers. A perturbation with wavenumbers $k_y$ and $k_z$ will create a peak at the one-dimensional wavenumber $k = \sqrt{k_y^2 + k_z^2}$.

\subsubsection{Calculation of growth speeds}

For each simulation, we have calculated a time-sequence of one-dimensional Fourier spectra. We use the central differential method to calculate the growth speed of perturbations in layer position in units of \mbox{$\text{pc}\,\text{Myr}^{-1}$}. This yields the growth speed as a function of time for each wavenumber. As we have multiple realizations of each simulation, we average growth speeds across these realizations, and calculate the standard deviation.

For monochromatic simulations, we are only interested in the growth speed for the wavenumber which was originally excited, averaged across the five realizations per wavenumber. We therefore extract the growth speed as a function of time for this single wavenumber. For square cross-section monochromatic simulations, we can directly extract the relevant wavenumber from the two-dimensional Fourier transform without using our circular averaging. We combine the results from all the monochromatic simulations at different wavenumbers to produce the growth speed as a function of time and as a function of all the excited wavenumbers.

\section{Results}

\begin{figure}
   \centering
   \includegraphics[width=\linewidth]{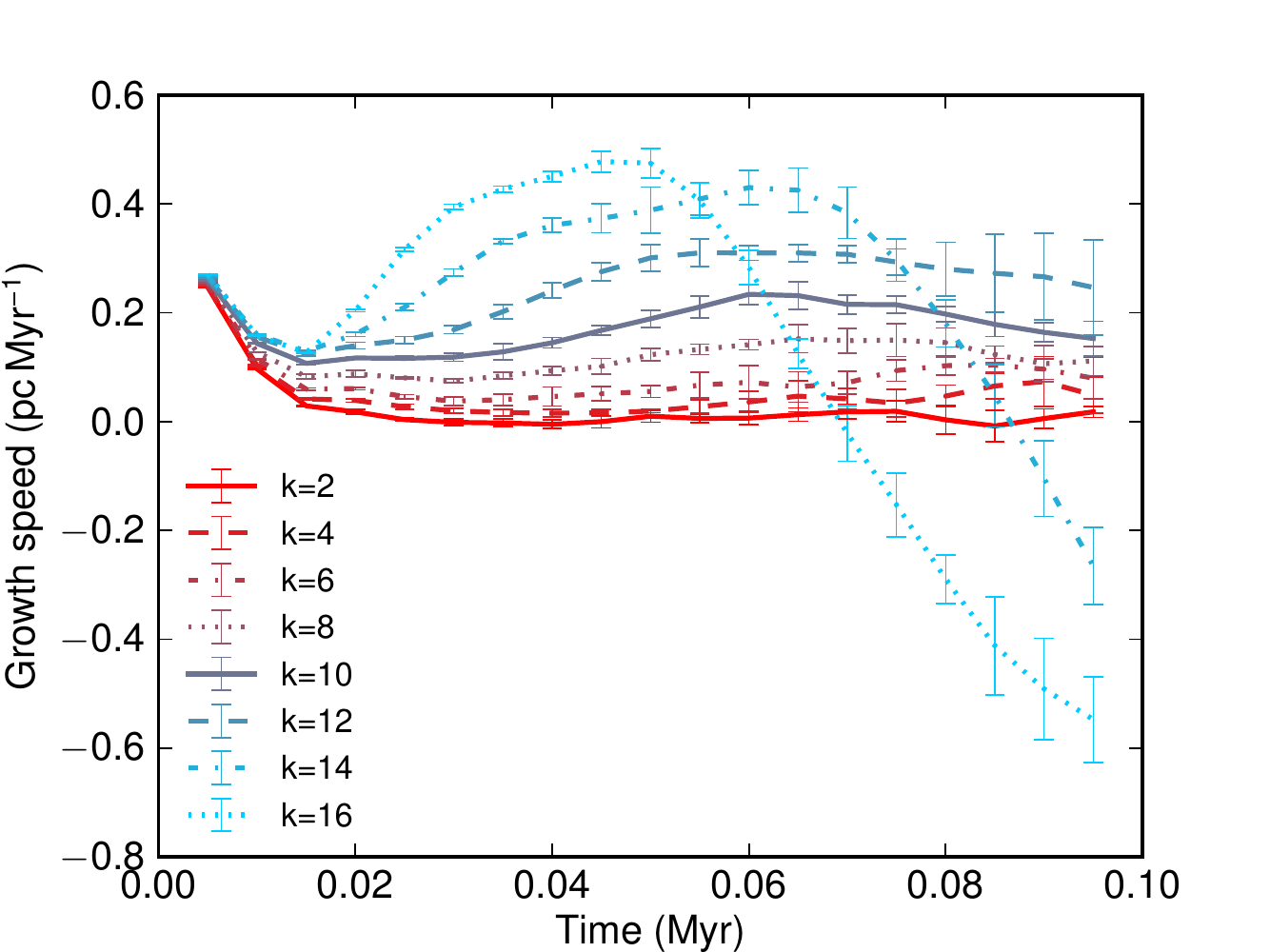}
   \caption[1D monochromatic growth speeds]{The growth speeds for one-dimensional monochromatic perturbations as a function of time. Each line represents a different wavenumber; error bars show the standard deviation across realizations.}
   \label{fig:1D_reverse_mono}
\end{figure}

The collision of the flows creates a shock-compressed layer. As expected, the thickness of this layer increases linearly with time, and the rate of increase does not depend strongly on the type of initial velocity perturbation. However, the rate does differ between narrow and square cross-section simulations. We fit the thickness as a function of time for each simulation using the linear function $\xi_\mathrm{FIT}(t) = \kappa \, t$, and average over the different simulations to find $\kappa = 0.47\,\text{pc}\,\text{Myr}^{-1}$ for narrow cross-section simulations and $\kappa = 0.22\,\text{pc}\,\text{Myr}^{-1}$ for square cross-section simulations. This is faster than the \mbox{$0.019\,\text{pc}\,\text{Myr}^{-1}$} predicted for an idealized isothermal collision.

This increase is not due to insufficient resolution, as the layer grows faster for the higher-resolution narrow cross-section simulations than the lower-resolution square cross-section simulations. It is due to the bending of the layer, which reduces the effective ram pressure. This causes the layer to reach a lower density than for the idealized isothermal case.

For a layer at an angle $\theta$, the component of the inflowing velocity $v_0$ that contributes to the ram pressure is $v_0 \cos \theta$. Since ram pressure is given by $\rho v^2$, the effective ram pressure and therefore the layer density is reduced by a factor of $\cos^2 \theta$. This increases the thickness of the layer by a factor of $\sec^2 \theta$. Both our initial perturbation and the NTSI create significant bending angles, which greatly increase the thickness of the layer.

This effect has been identified previously in simulations of the NTSI. \citet{BlondinMarks1996} found their layers grew faster than expected even in the absence of the NTSI due to the growth of the Kelvin--Helmholtz instability. In simulations of cold dense layers confined by supersonic shocks, \citet{FoliniWalder2006} found that all their layers reached a maximum density contrast relative to the inflowing medium of $\sim 30$ regardless of the Mach number. This is similar to our layers, which have average density contrasts of approximately 16 and 35 for the narrow and square cross-section simulations, respectively.

Figure~\ref{fig:1D_reverse_mono} shows the growth speed of different modes as a function of time for simulations with one-dimensional monochromatic perturbations. The results for each mode are taken as the average from a set of simulations where only that mode is perturbed. As expected, the growth speeds of all wavenumbers begin at a value corresponding to the amplitude of the initial perturbation, and then quickly decay.

Following the decay of the initial perturbation, the growth speeds of perturbations at small wavenumbers do not increase appreciably during the simulation time. However, higher wavenumber perturbations start to grow, and reach larger growth speeds than the initial perturbation, before decreasing again at later times. This increase is due to the NTSI; the later decline is due to the layer becoming too thick to support perturbations of higher wavenumbers.

We can compare the growth of the NTSI to the maximum and minimum wavelength limits defined by Equations~\ref{eq:NTSI_minimum_wavelength} and~\ref{eq:NTSI_maximum_wavelength}, respectively, and the corresponding minimum wavenumber limit
\begin{equation}
   \label{eq:NTSI_minimum_wavenumber}
   k \gg \frac{1}{c_\mathrm{s} t}\;,
\end{equation}
and the maximum resolvable wavenumber
\begin{equation}
   \label{eq:NTSI_maximum_wavenumber}
   k \lesssim \frac{1}{2 \, \xi_\mathrm{FIT}(t)}\;.
\end{equation}

We find the minimum wavenumber limit is not reliable. At the end of the simulation time the minimum dimensionless wavenumber should be much greater than 26, and higher at earlier times. However, Figure~\ref{fig:1D_reverse_mono} suggests a minimum wavenumber closer to 10. The minimum wavenumber limit assumes that wavenumbers do not grow in less than the sound crossing time across a wavelength. As described in Section~\ref{sec:growth_to_saturation}, combining this with the limit of small bending angles also implies that the growth speeds of perturbations should remain at most \text{weakly supersonic}. We note that several of these modes achieve growth speeds which imply supersonic velocities at the peaks of the perturbation, but none reach more than approximately Mach 2.5.

However, we do confirm the maximum wavenumber limit, equation~\ref{eq:NTSI_maximum_wavenumber}. We use the averaged rate of layer thickness growth described above, with a different value for narrow and square cross-section simulations, to estimate the layer thickness and the corresponding maximum resolvable wavenumber at each time. In our simulations, growth at each wavenumber stops and decays close to the predicted value.

\subsection{Monochromatic perturbations}

For monochromatic perturbations, we first consider simulations with a one-dimensional perturbation where we excite a single wavenumber using the initial velocity perturbation given by equation~\ref{eq:IC_1D_vel_pert}. We then consider simulations with a two-dimensional perturbation, where we excite a single set of wavenumbers using the velocity perturbation given by equation~\ref{eq:IC_2D_vel_pert}. As noted in Section~\ref{sec:monochromatic_perturbations}, we cannot excite exactly the same wavenumbers in these simulations.

\subsubsection{One-dimensional perturbations}

\begin{figure}
   \centering
   \includegraphics[width=\linewidth]{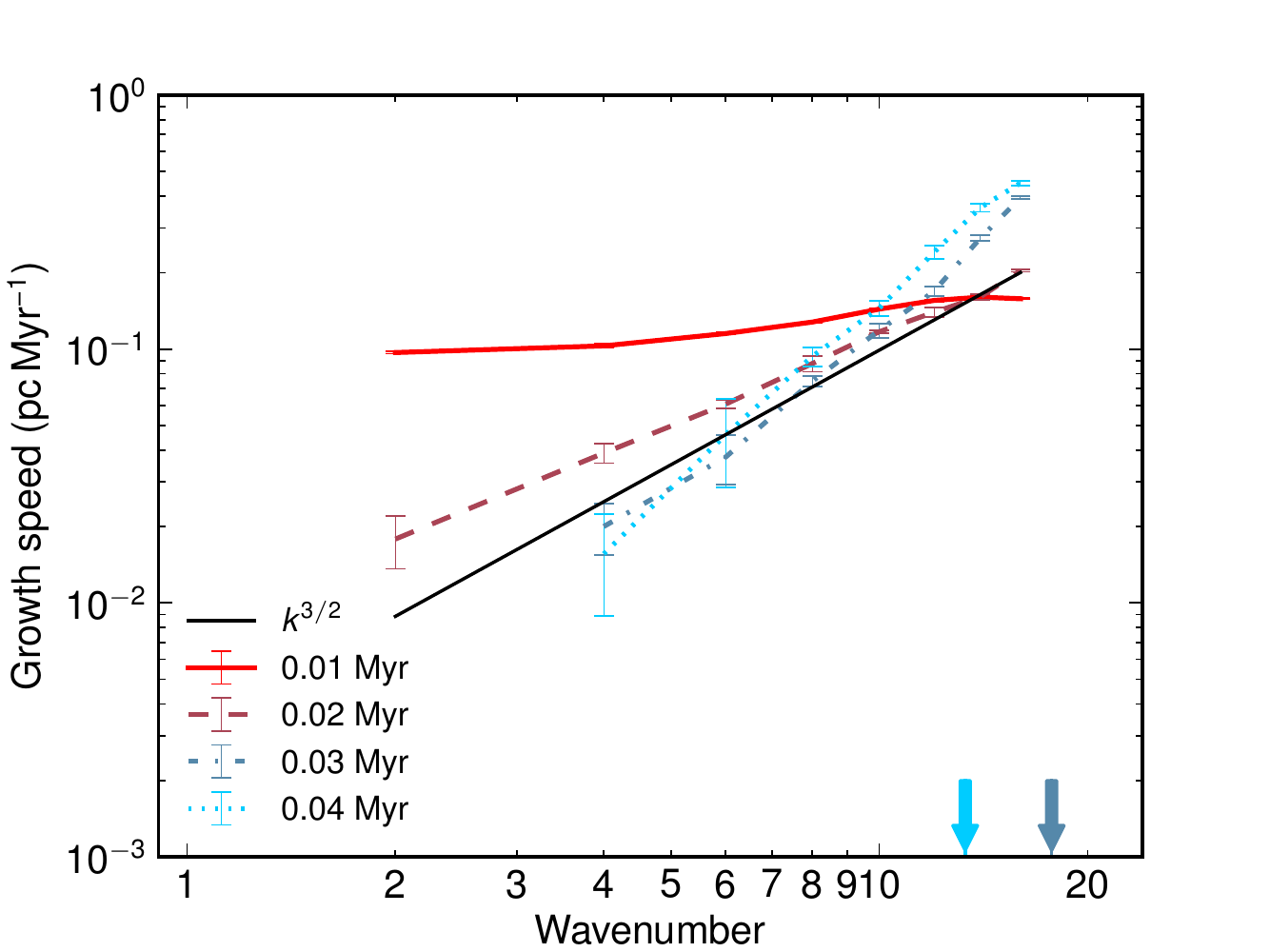}
   \includegraphics[width=\linewidth]{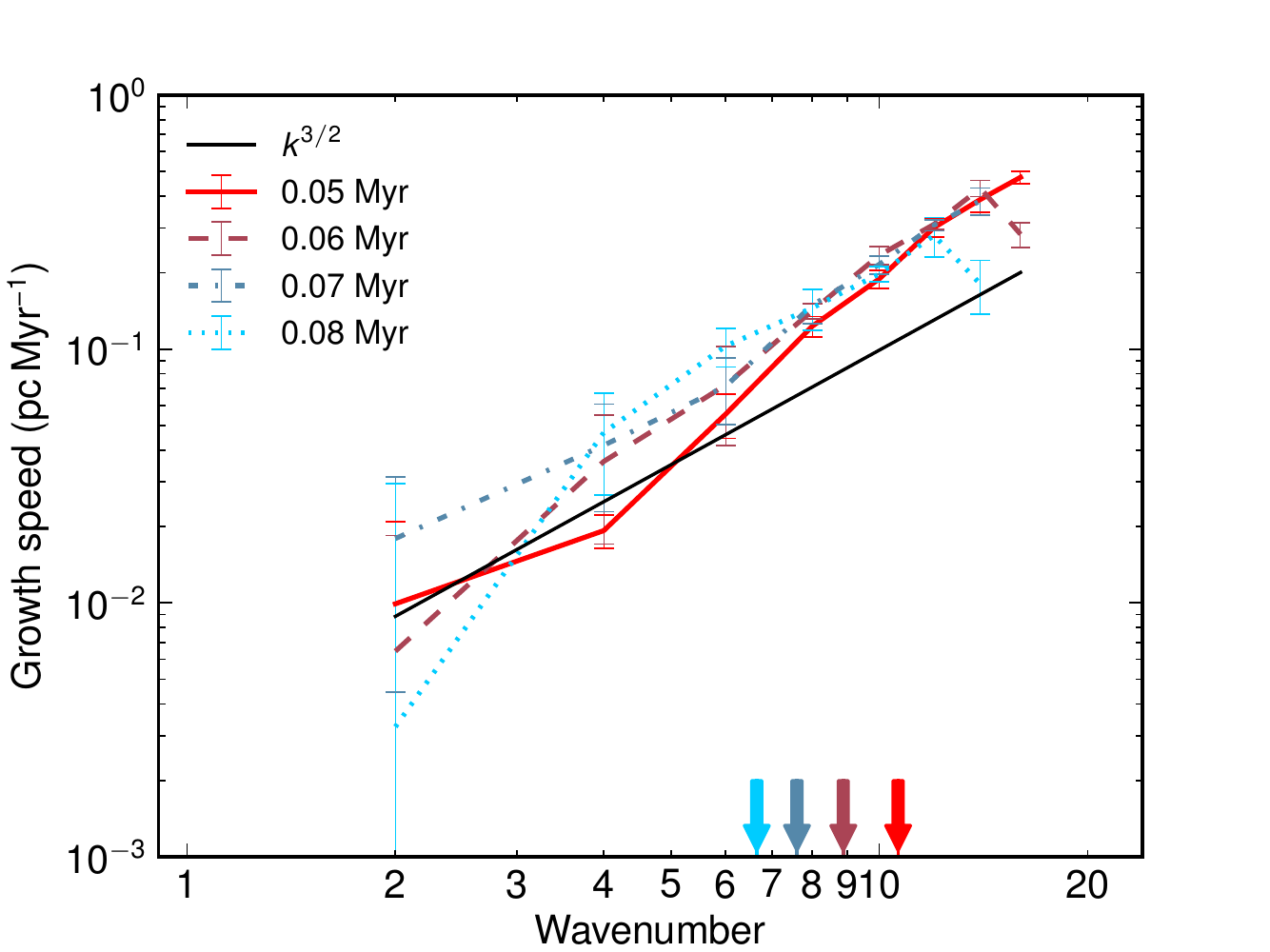}
   \caption[One-dimensional monochromatic perturbation growth speeds]{The growth speeds for one-dimensional monochromatic perturbations as a function of wavenumber. Each line represents results from a different timestep; the upper frame displays the growth speed at early times and the lower frame displays the growth speeds at later times. Error bars show the standard deviation across realizations. The gradient of the solid black line indicates the relation predicted by \citet{Vishniac1994}. The arrows indicate the maximum resolvable wavenumber for the correspondingly coloured timestep; arrows run from latest to earliest timestep going left to right, and some arrows may be at higher wavenumbers than can be plotted.}
   \label{fig:1D_mono}
\end{figure}

\begin{figure}
   \centering
   \includegraphics[width=\linewidth]{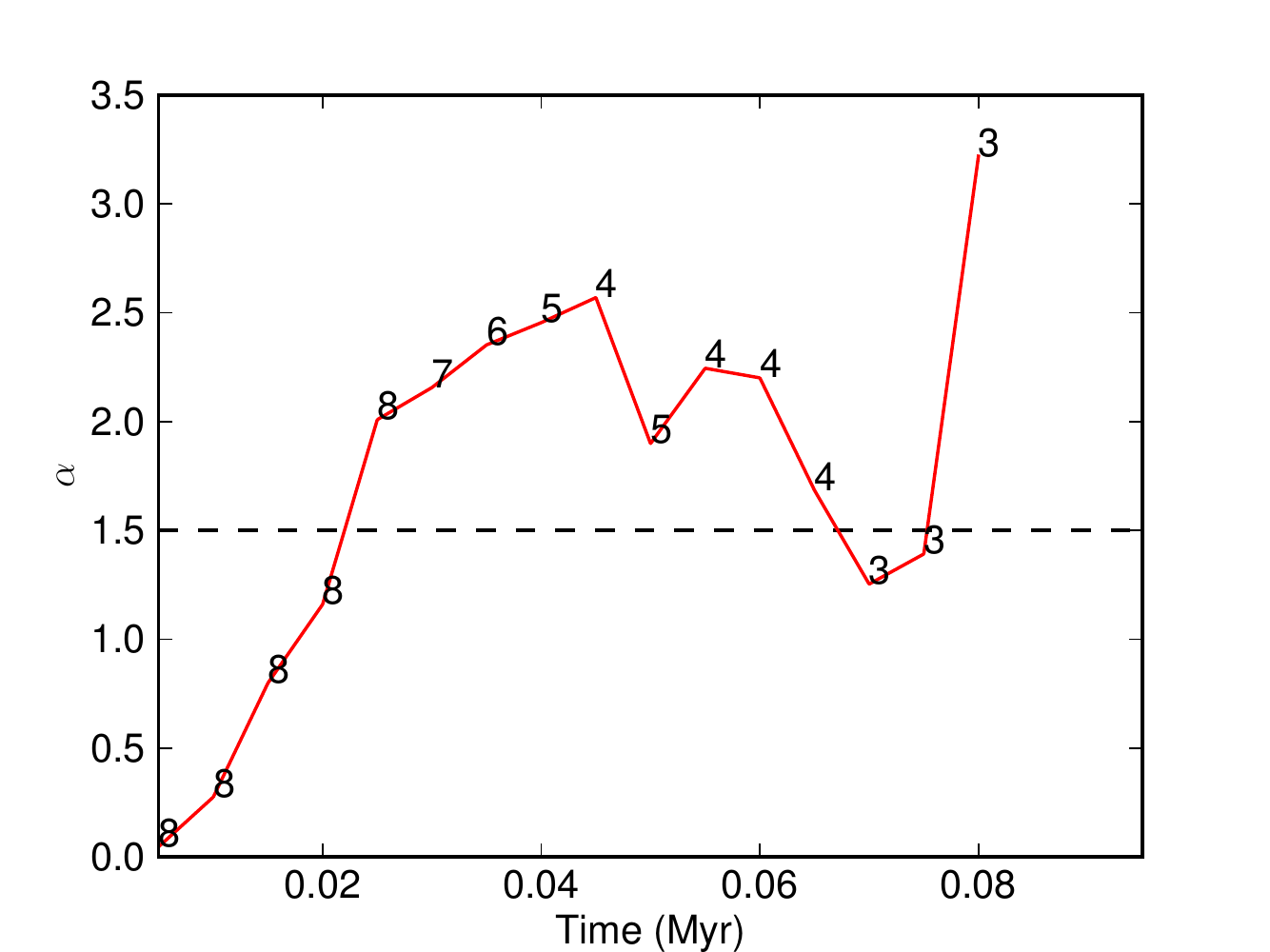}
   \caption[One-dimensional monochromatic perturbation fits]{Indices of power-law fits to growth speeds of one-dimensional monochromatic perturbations as a function of time. The number of points available to construct each fit is also shown; fits constructed with fewer points are less reliable. The black dashed line indicates the prediction of \citet{Vishniac1994}.}
   \label{fig:1D_mono_fits}
\end{figure}

Figure~\ref{fig:1D_mono} shows the growth speeds for simulations with one-dimensional monochromatic perturbations. As expected, at early times the slope of the line is roughly flat, as all modes are given the same initial perturbation amplitude. At later times, higher wavenumbers grow faster than smaller wavenumbers, appearing to approximately match the slope predicted by \citet{Vishniac1994} of $k^{1.5}$. The highest wavenumbers begin to decay at the latest times, but at somewhat higher wavenumbers than predicted by equation~\ref{eq:NTSI_maximum_wavenumber}.

Figure~\ref{fig:1D_mono_fits} shows the index of power-law fits to the growth speeds as a function of wavenumber. We fit from the smallest wavenumbers to the maximum resolvable wavenumber at each time. As expected, the index begins at approximately zero, but rises rapidly. This rise brings it somewhat above the predicted index of 1.5, instead lying in the range of approximately 2.0 to 2.5.

\subsubsection{Two-dimensional perturbations}

\begin{figure}
   \centering
   \includegraphics[width=\linewidth]{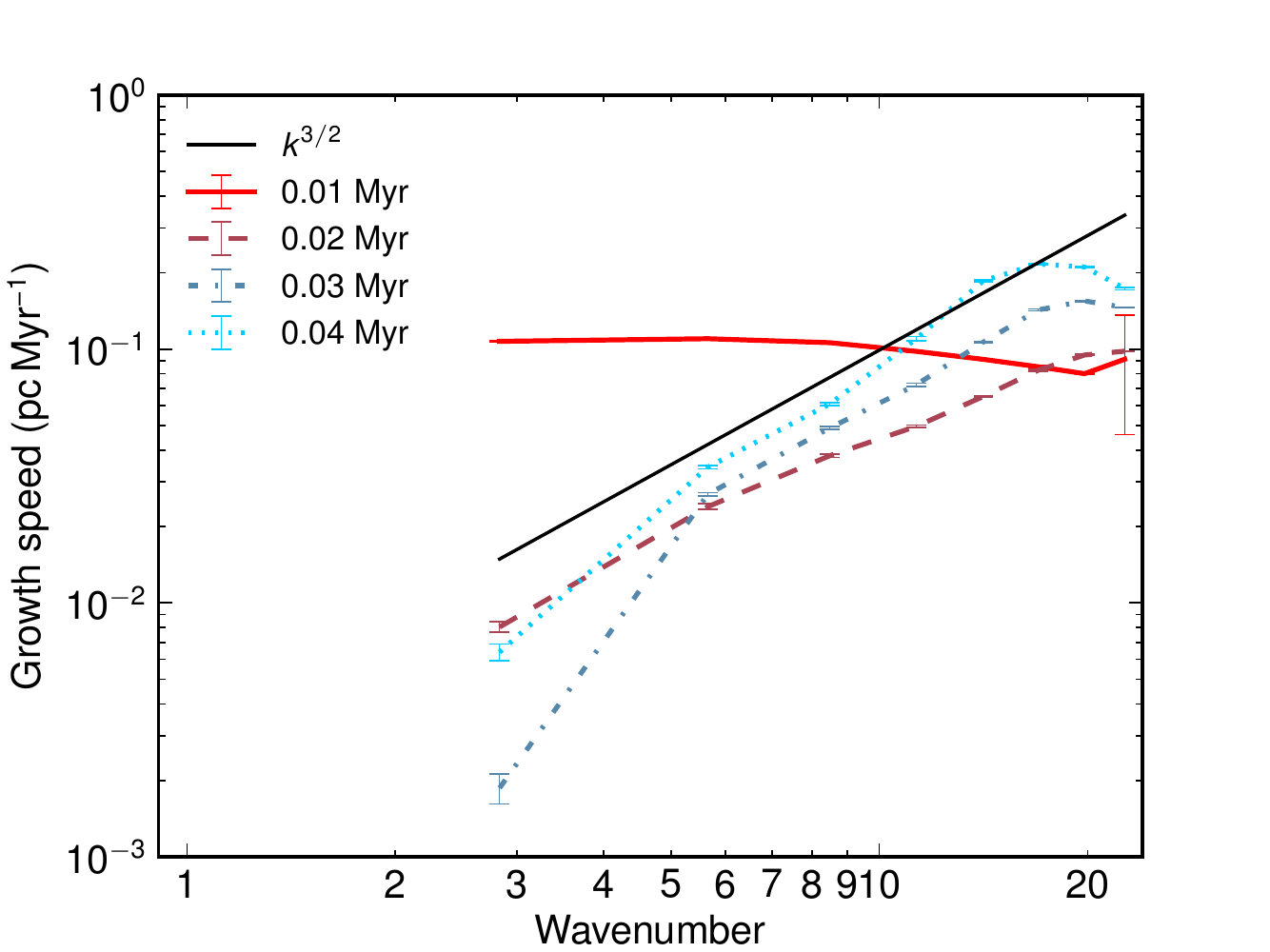}
   \includegraphics[width=\linewidth]{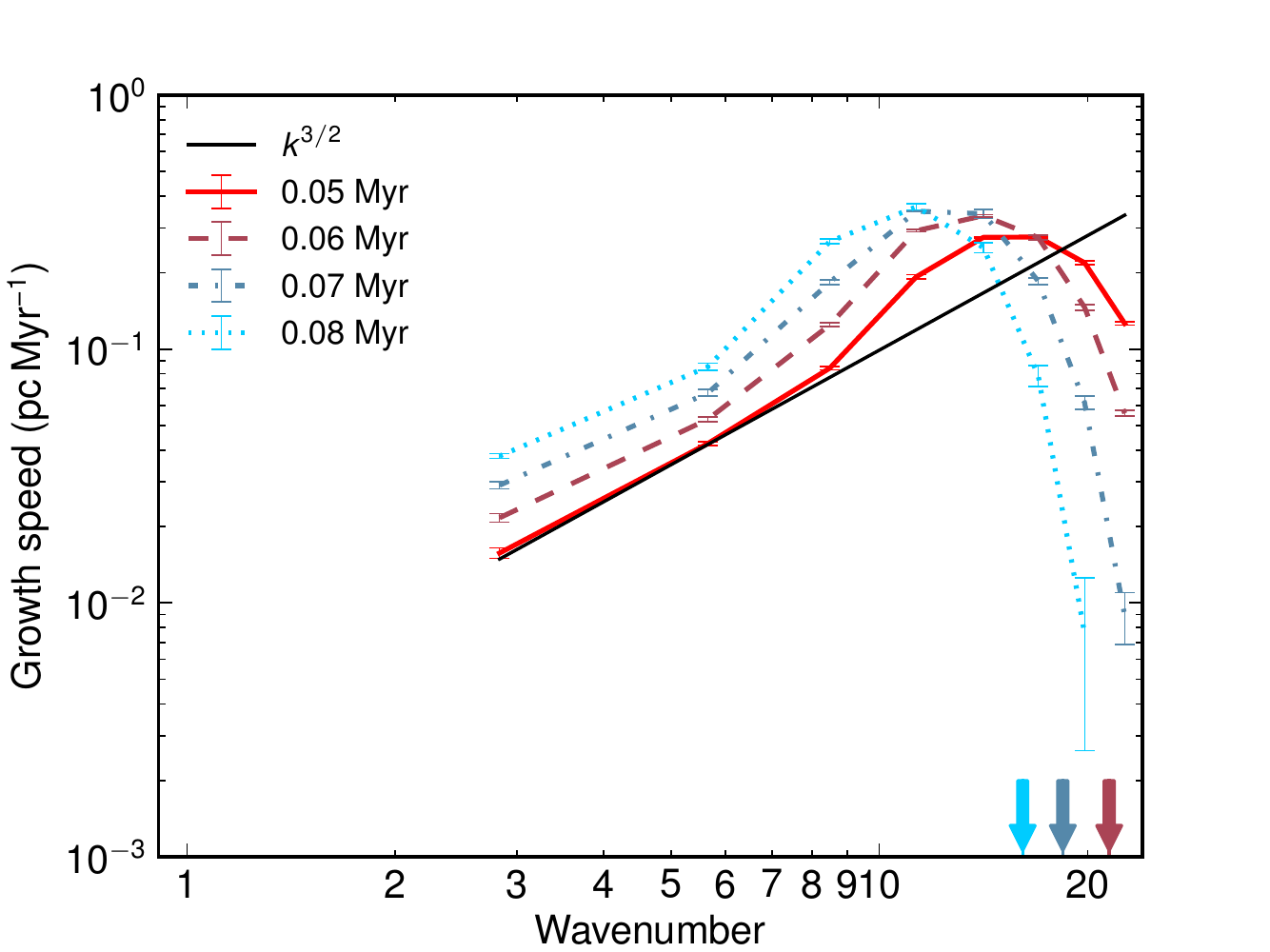}
   \caption[Two-dimensional monochromatic perturbation growth speeds]{The growth speeds from simulations with two-dimensional monochromatic perturbations as a function of wavenumber, as for Figure~\ref{fig:1D_mono}.}
   \label{fig:2D_mono}
\end{figure}

\begin{figure}
   \centering
   \includegraphics[width=\linewidth]{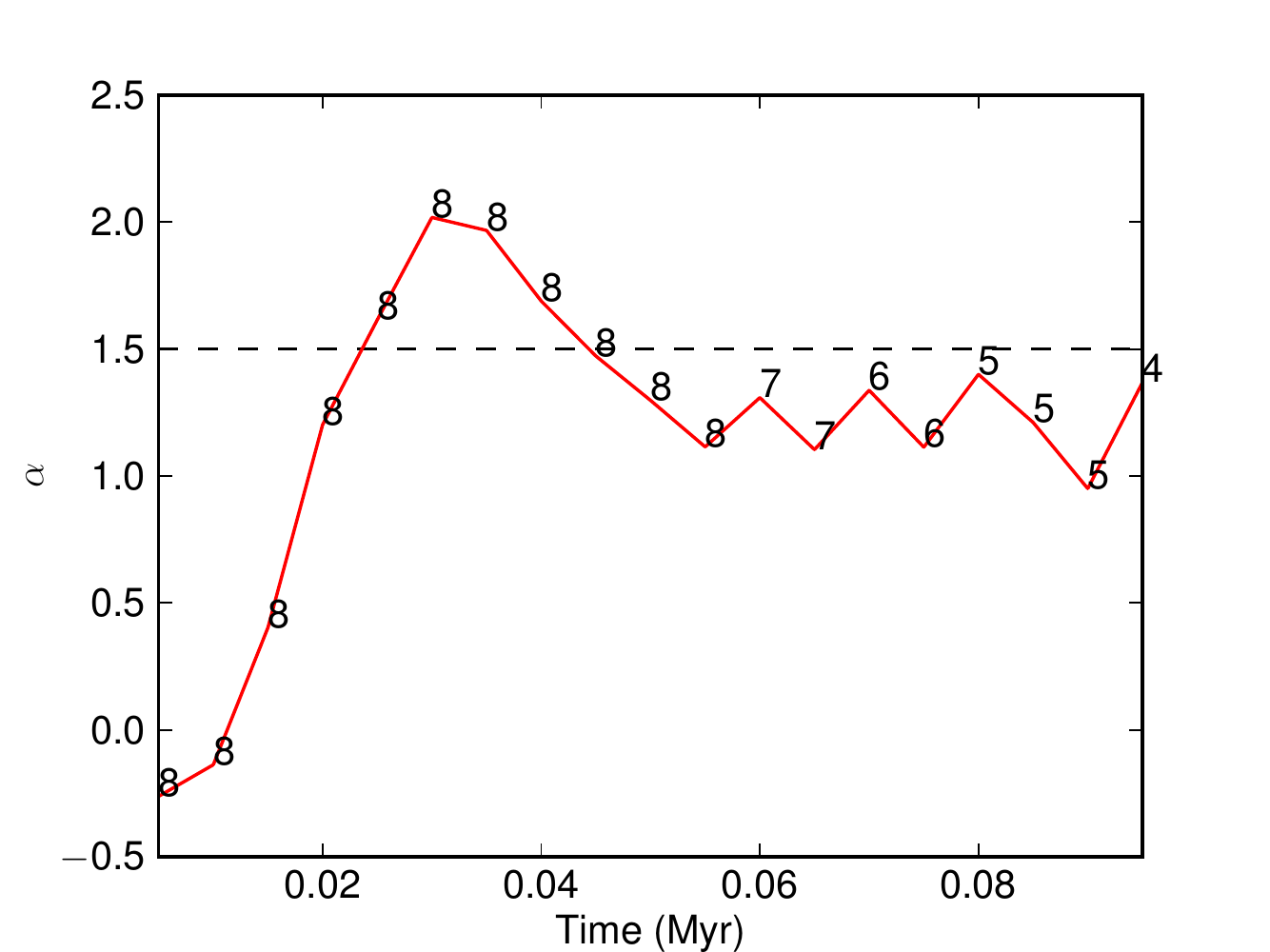}
   \caption[Two-dimensional monochromatic perturbation fits]{Indices of power-law fits to growth speeds from simulations with two-dimensional monochromatic perturbations, as for Figure~\ref{fig:1D_mono_fits}.}
   \label{fig:2D_mono_fits}
\end{figure}

Figure~\ref{fig:2D_mono} shows the growth speeds for simulations with two-dimensional monochromatic perturbations. Again the slopes begin fairly flat before rising to approximately match the prediction of \citeauthor{Vishniac1994}. At later times the decay of higher wavenumbers is evident and occurs at approximately the wavenumber predicted by equation~\ref{eq:NTSI_maximum_wavenumber}.

Figure~\ref{fig:2D_mono_fits} shows the index of power-law fits to the growth speeds as a function of wavenumber. This initially rises from zero, as expected, and then remains close to the predicted value of 1.5.

\subsection{White-noise perturbations}

White-noise perturbations are created by a superposition of velocity perturbations of equal amplitude and random phase, as described in Section~\ref{sec:white_noise}. Not all modes can grow in any given simulation; some modes will grow while others will decay, creating the large standard deviations shown in Figures~\ref{fig:1D_mix} and~\ref{fig:2D_mix}.

Unlike for monochromatic perturbations, where we consider only a single wavenumber from each simulation, we consider the growth speed of a range of wavenumbers and wavevectors from each simulation. For simulations with two-dimensional perturbations, this means we have to perform circular averaging, as described in Section~\ref{sec:fourier_transform}, to produce the growth speed as a function of the one-dimensional wavenumber.

\subsubsection{One-dimensional perturbations}

\begin{figure}
   \centering
   \includegraphics[width=\linewidth]{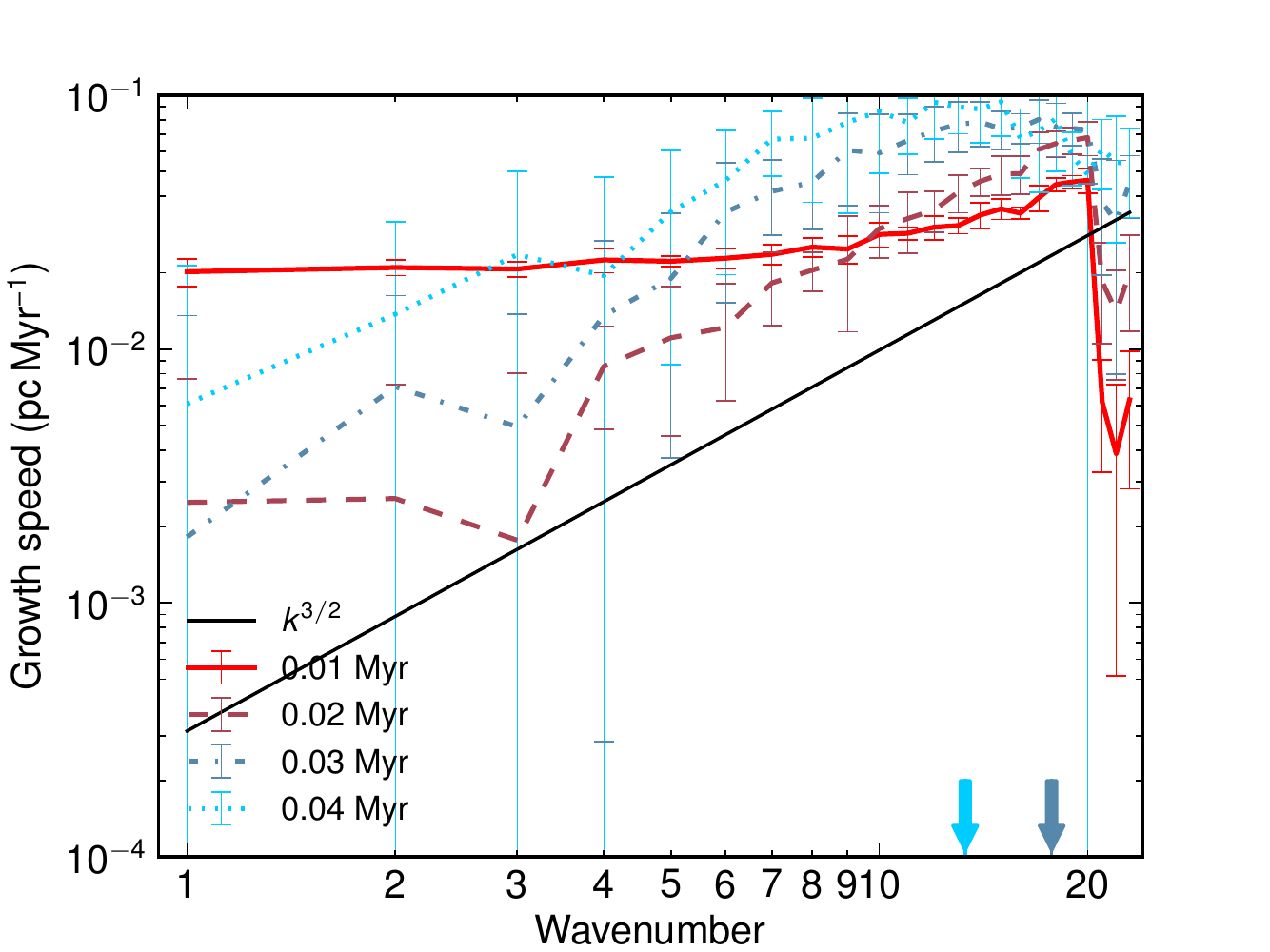}
   \includegraphics[width=\linewidth]{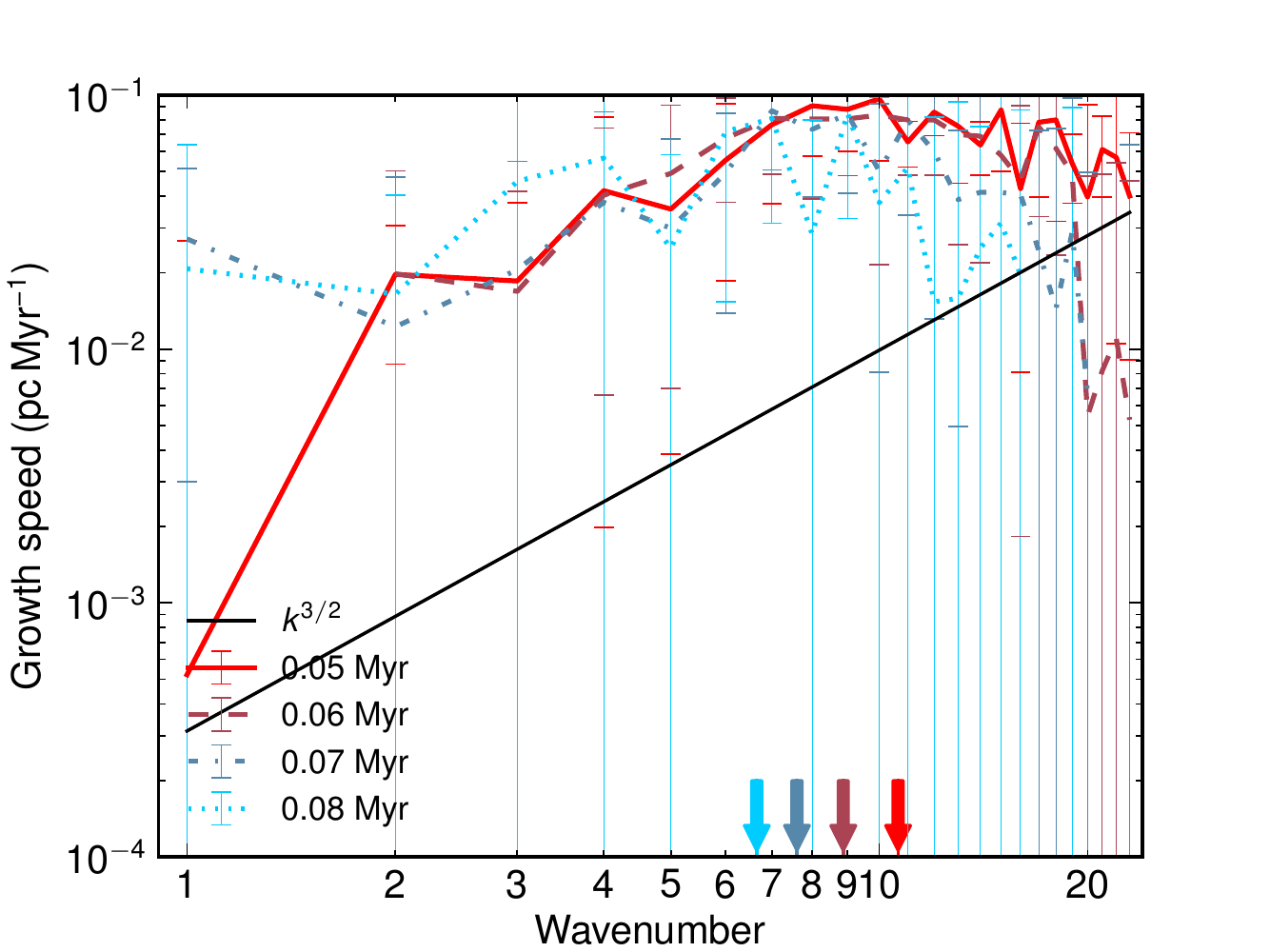}
   \caption[One-dimensional white-noise growth speeds]{The growth speeds from simulations with one-dimensional white-noise perturbations as a function of wavenumber, as for Figure~\ref{fig:1D_mono}.}
   \label{fig:1D_mix}
\end{figure}

\begin{figure}
   \centering
   \includegraphics[width=\linewidth]{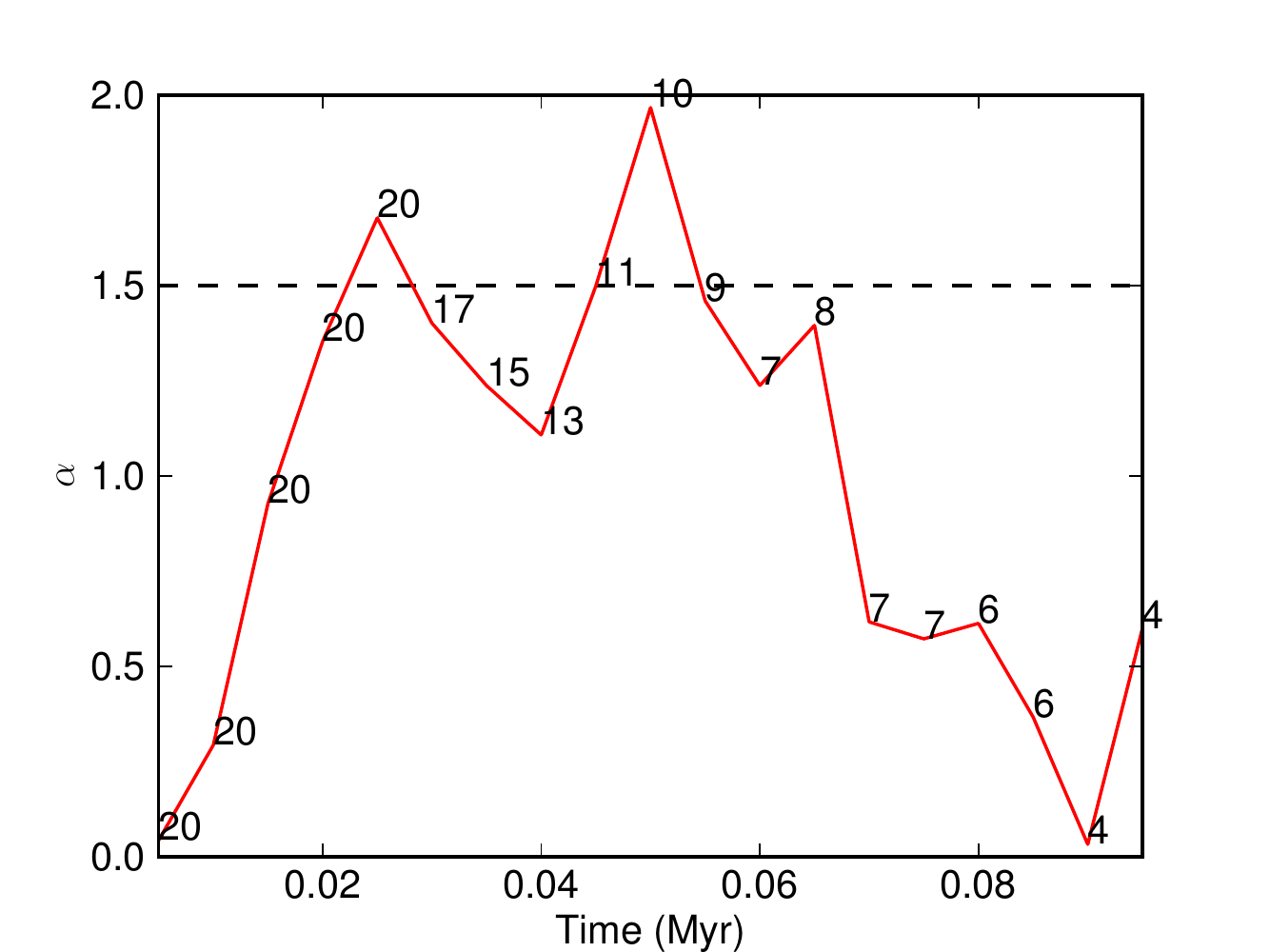}
   \caption[One-dimensinonal white-noise fits]{Indices of power-law fits to growth speeds from simulations with one-dimensional white-noise perturbations, as for Figure~\ref{fig:1D_mono_fits}.}
   \label{fig:1D_mix_fits}
\end{figure}

Figure~\ref{fig:1D_mix} shows the growth speeds for simulations with one-dimensional white-noise perturbations. Again the slopes begin fairly flat before rising to approximately match the prediction of \citeauthor{Vishniac1994}. At later times, the decay of higher wavenumbers is evident and occurs at or slightly above the wavenumber predicted by equation~\ref{eq:NTSI_maximum_wavenumber}.

Figure~\ref{fig:1D_mix_fits} shows the index of power-law fits to the growth speeds as a function of wavenumber. As expected this initially rises from zero and then remains close to the predicted value of 1.5, although with considerable scatter. The results at later times should be ignored due to the low number of wavenumbers that can still be resolved as the layer grows in thickness.

\subsubsection{Two-dimensional perturbations}

\begin{figure}
   \centering
   \includegraphics[width=\linewidth]{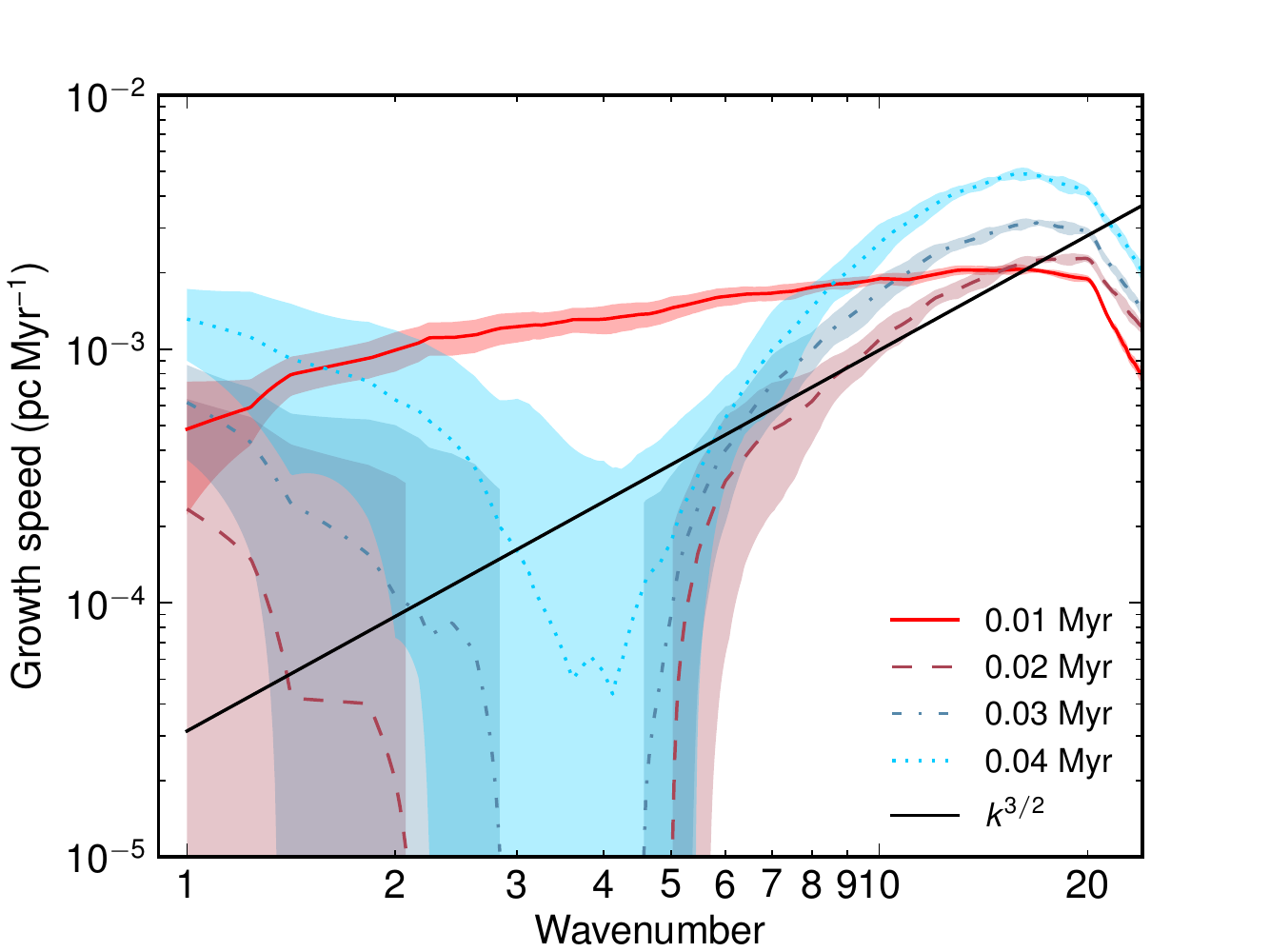}
   \includegraphics[width=\linewidth]{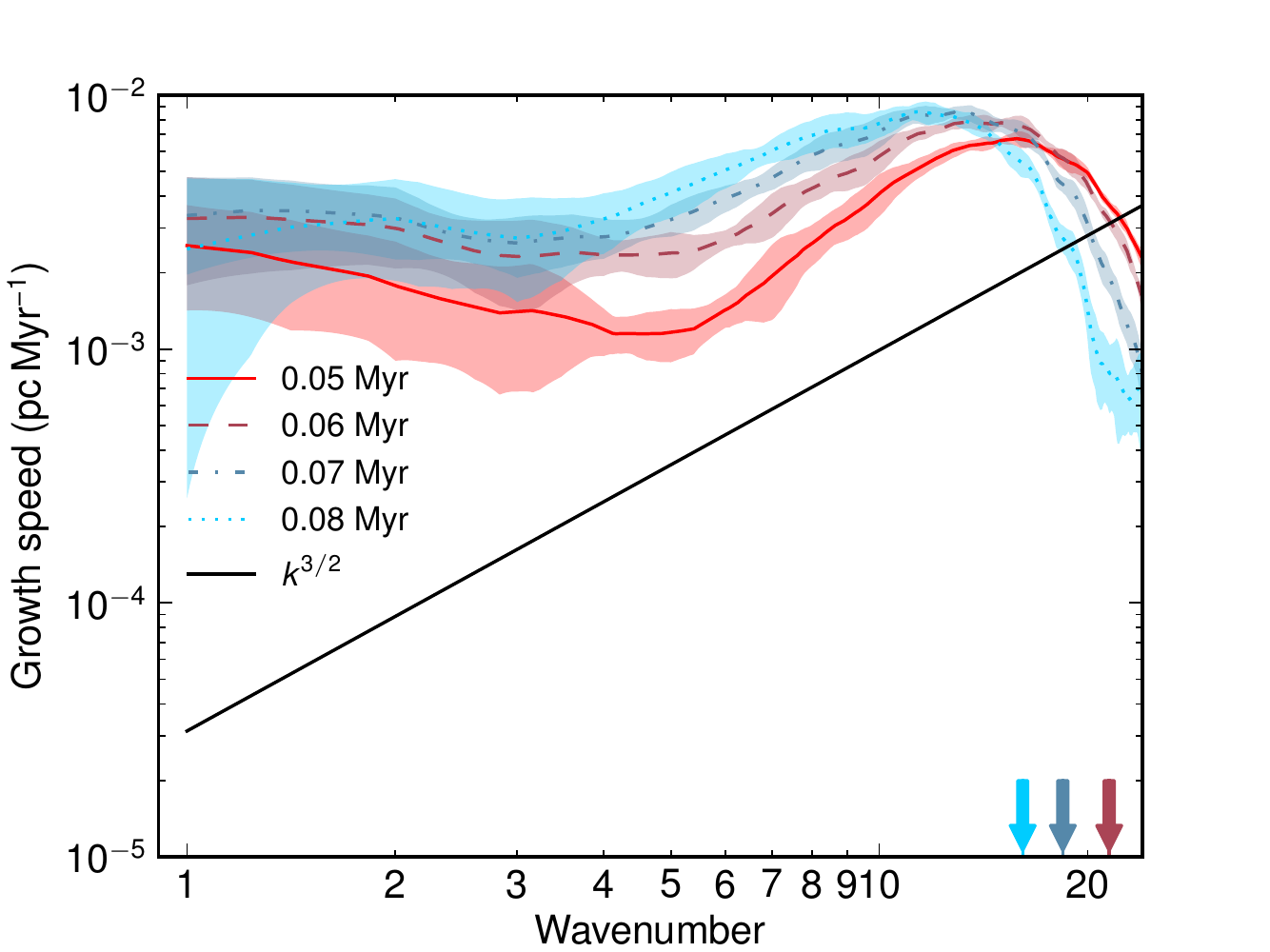}
   \caption[Two-dimensional white-noise growth speeds]{The growth speeds from simulations with two-dimensional white-noise perturbations as a function of wavenumber, as for Figure~\ref{fig:1D_mono}.}
   \label{fig:2D_mix}
\end{figure}

\begin{figure}
   \centering
   \includegraphics[width=\linewidth]{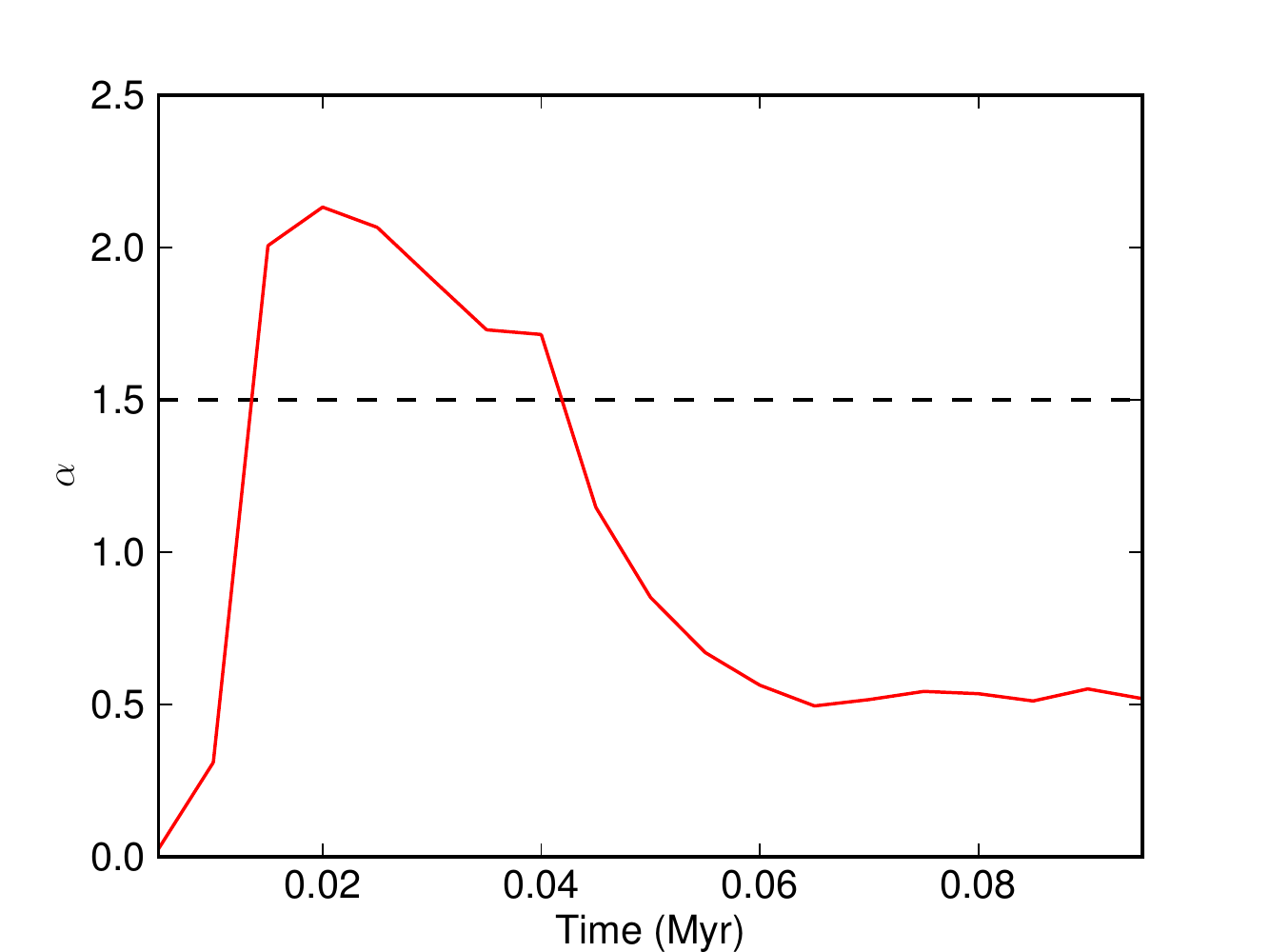}
   \caption[Two-dimensional white-noise fits]{Indices of power-law fits to growth speeds from simulations with two-dimensional white-noise perturbations, as for Figure~\ref{fig:1D_mono_fits}.}
   \label{fig:2D_mix_fits}
\end{figure}

Figure~\ref{fig:2D_mix} shows the growth speeds for simulations with two-dimensional white-noise perturbations. Due to the circular averaging, results are presented as a continuous line with a shaded error zone instead of individual points. These growth speeds are more complex than for the simulations with monochromatic perturbations. It is clear that higher wavenumbers generally grow faster up to some limiting wavenumber. However, there is a large decrease in growth speeds at smaller wavenumbers at early times, followed by a growth at small wavenumbers at later times. Growth speeds appear to decay at wavenumbers somewhat smaller than predicted by equation~\ref{eq:NTSI_maximum_wavenumber}.

Figure~\ref{fig:2D_mix_fits} shows the index of power-law fits to the growth speeds as a function of wavenumber. This initially rises from zero as expected, before reaching approximately 2. The index then decreases with time, falling to approximately 0.7. Our power-law fits may not accurately reflect the growth of the NTSI for two reasons: the unexpected growth at small wavenumbers and the decay at larger wavenumbers below the maximum wavenumber predicted by equation~\ref{eq:NTSI_maximum_wavenumber}.

\subsection{Turbulence}

We now consider simulations that do not have an imposed sinusoidal velocity perturbation, but instead have a turbulent velocity field imposed on all the gas in the simulation. We use only square cross-section simulations, with either subsonic or supersonic turbulence as described in Section~\ref{sec:turbulence}. Although the turbulence itself is three-dimensional, it provides a time-varying two-dimensional set of perturbations as it is accreted onto the layer.

When fitting power-law slopes to the growth speeds, we fit from a minimum wavenumber of $k=8$ to help separate the effect of the turbulence and the NTSI. We fit to the maximum resolvable wavenumber predicted by equation~\ref{eq:NTSI_maximum_wavenumber}, or $k=64$ if this is smaller.

\subsubsection{Supersonic turbulence}

\begin{figure}
   \centering
   \includegraphics[width=\linewidth]{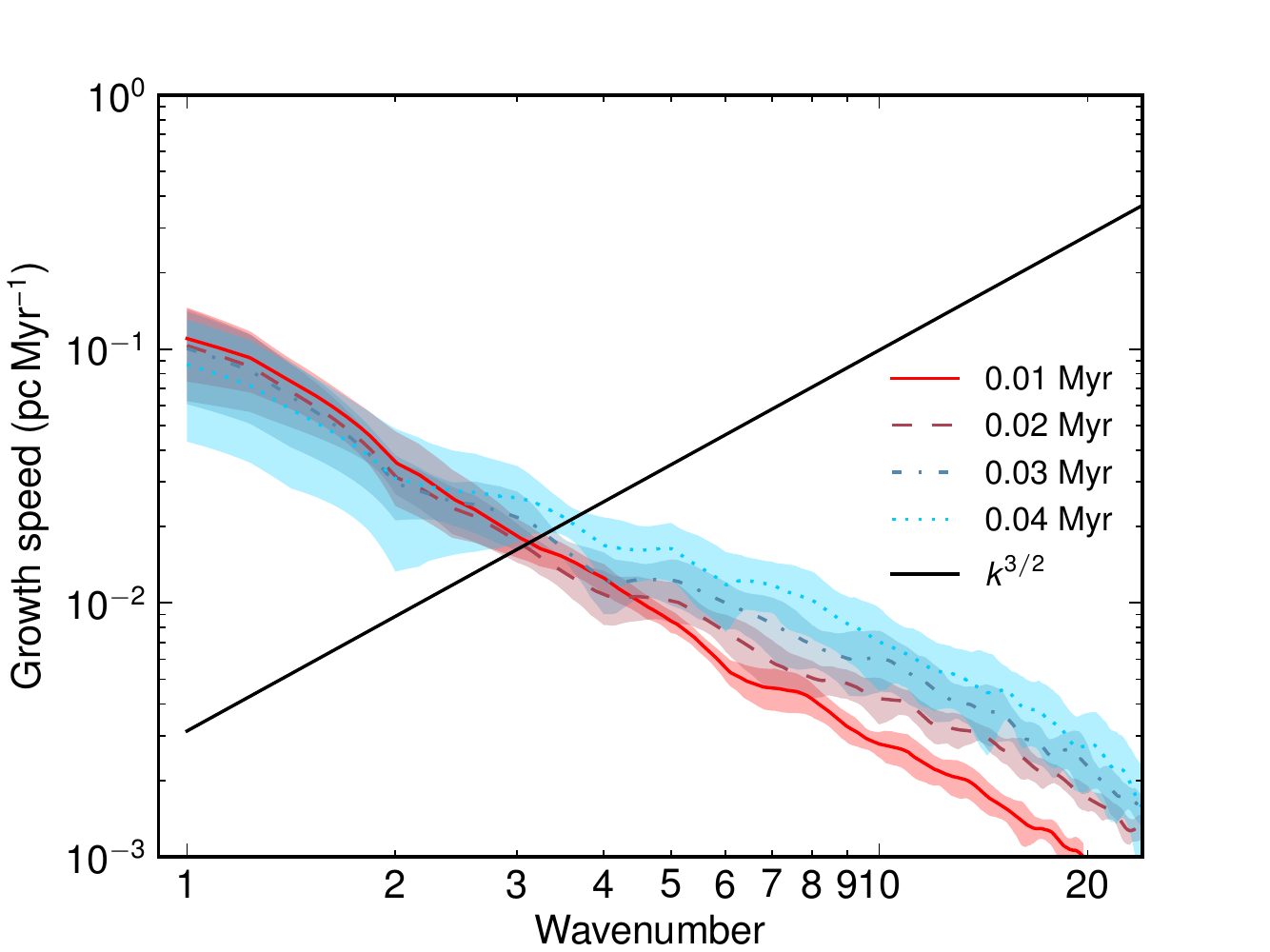}
   \includegraphics[width=\linewidth]{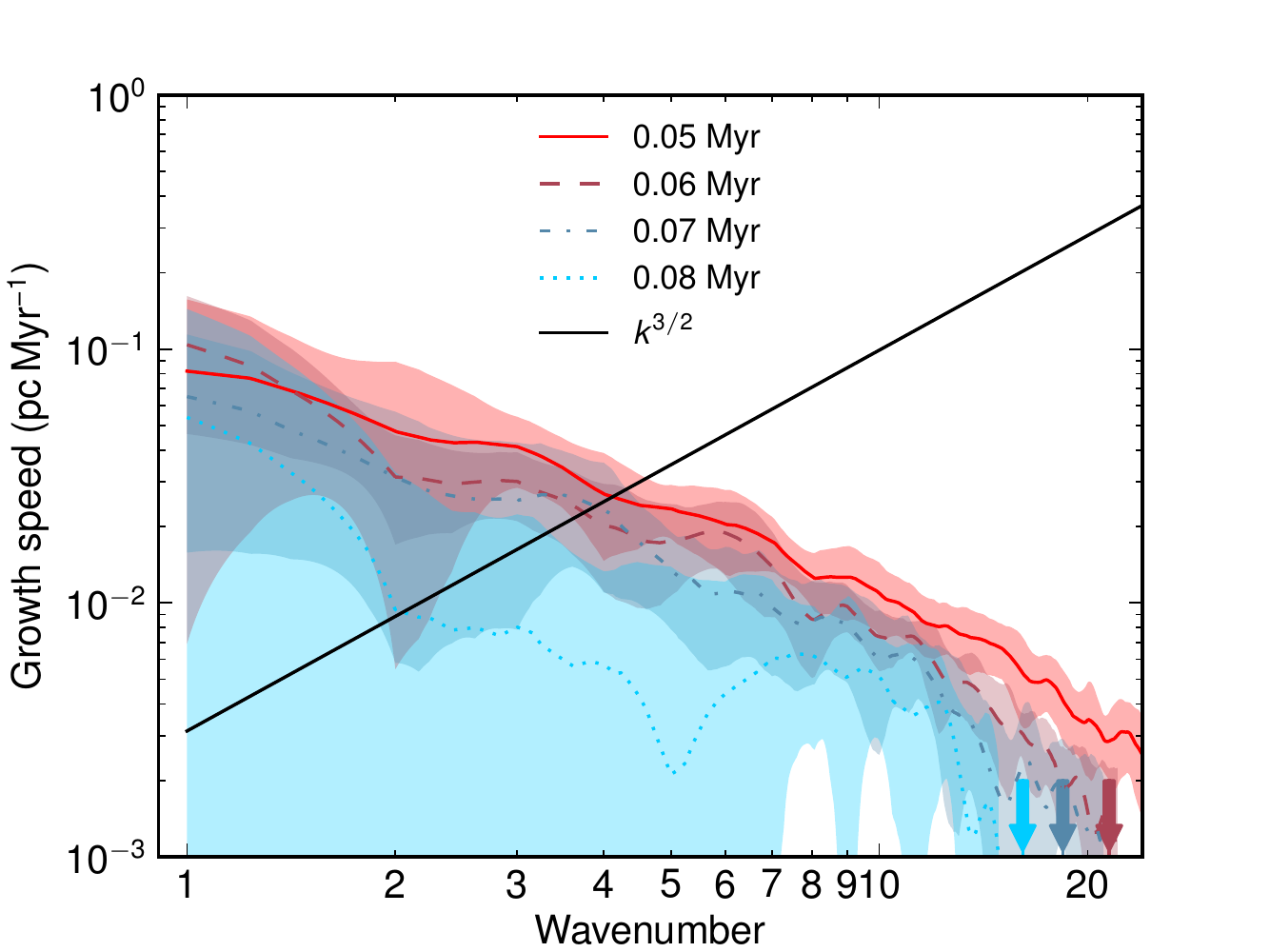}
   \caption[Supersonic turbulence growth speeds]{The growth speeds for simulations with supersonic turbulence as a function of wavenumber, as for Figure~\ref{fig:1D_mono}.}
   \label{fig:2D_turb}
\end{figure}

\begin{figure}
   \centering
   \includegraphics[width=\linewidth]{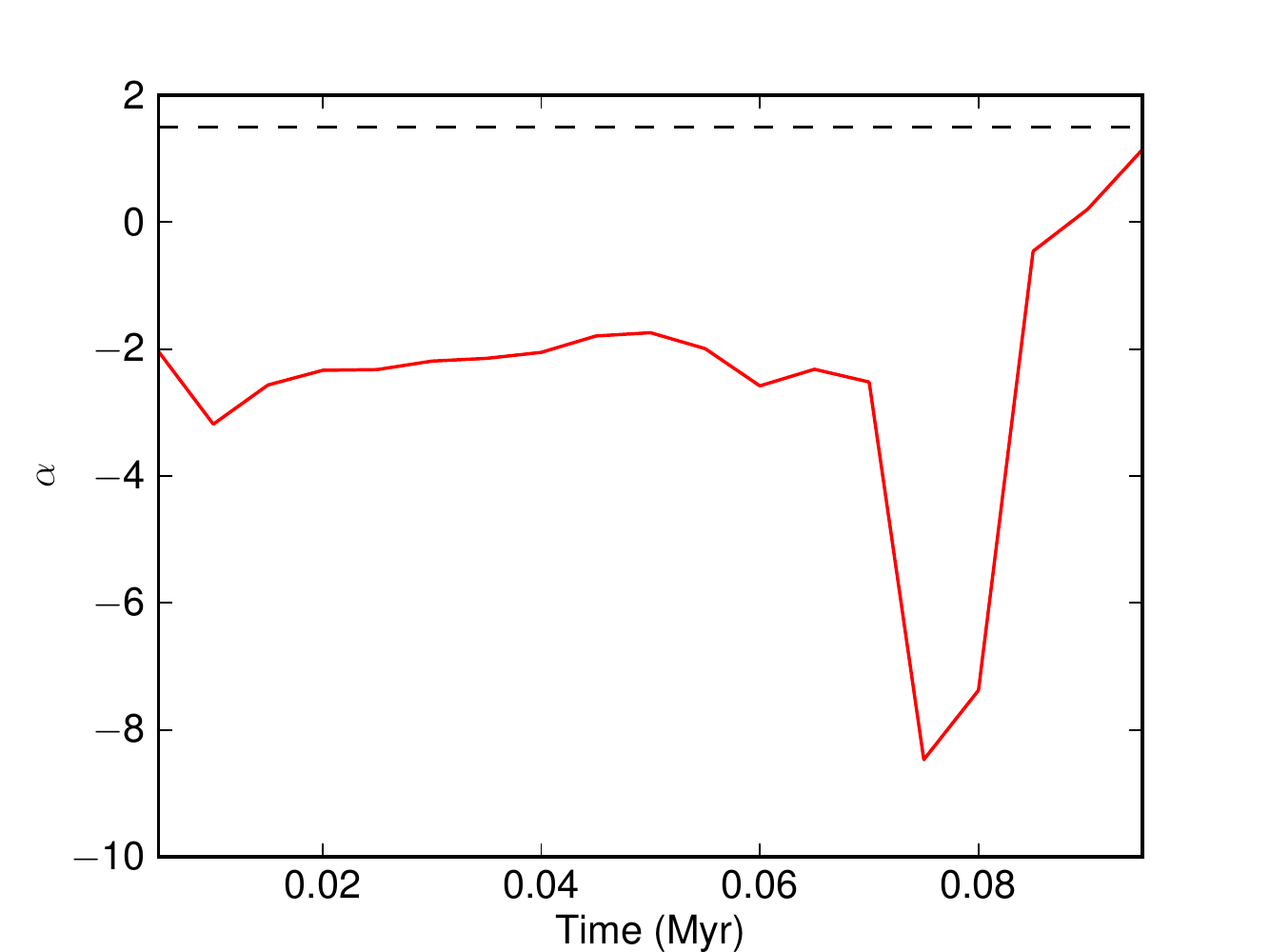}
   \caption[Supersonic turbulence fits]{Indices of power-law fits to growth speeds from simulations with supersonic turbulence, as for Figure~\ref{fig:1D_mono_fits}.}
   \label{fig:2D_turb_fits}
\end{figure}

Figure~\ref{fig:2D_turb} shows the growth speeds for simulations with supersonic turbulence. The influence of the turbulence, which is strongest at small wavenumbers, dominates the growth speeds. No evidence for the NTSI can be seen, with growth speeds decreasing with increasing wavenumber.

Figure~\ref{fig:2D_turb_fits} shows the index of power-law fits to the growth speeds as a function of wavenumber. The index begins strongly negative, in contrast to earlier simulations where the index begins at approximately zero. This is because there is not equal amplitude at all modes; the turbulence is stronger at smaller wavenumbers. The index remains at approximately $-2$. There is thus no evidence for the NTSI.

\subsubsection{Subsonic turbulence}

\begin{figure}
   \centering
   \includegraphics[width=\linewidth]{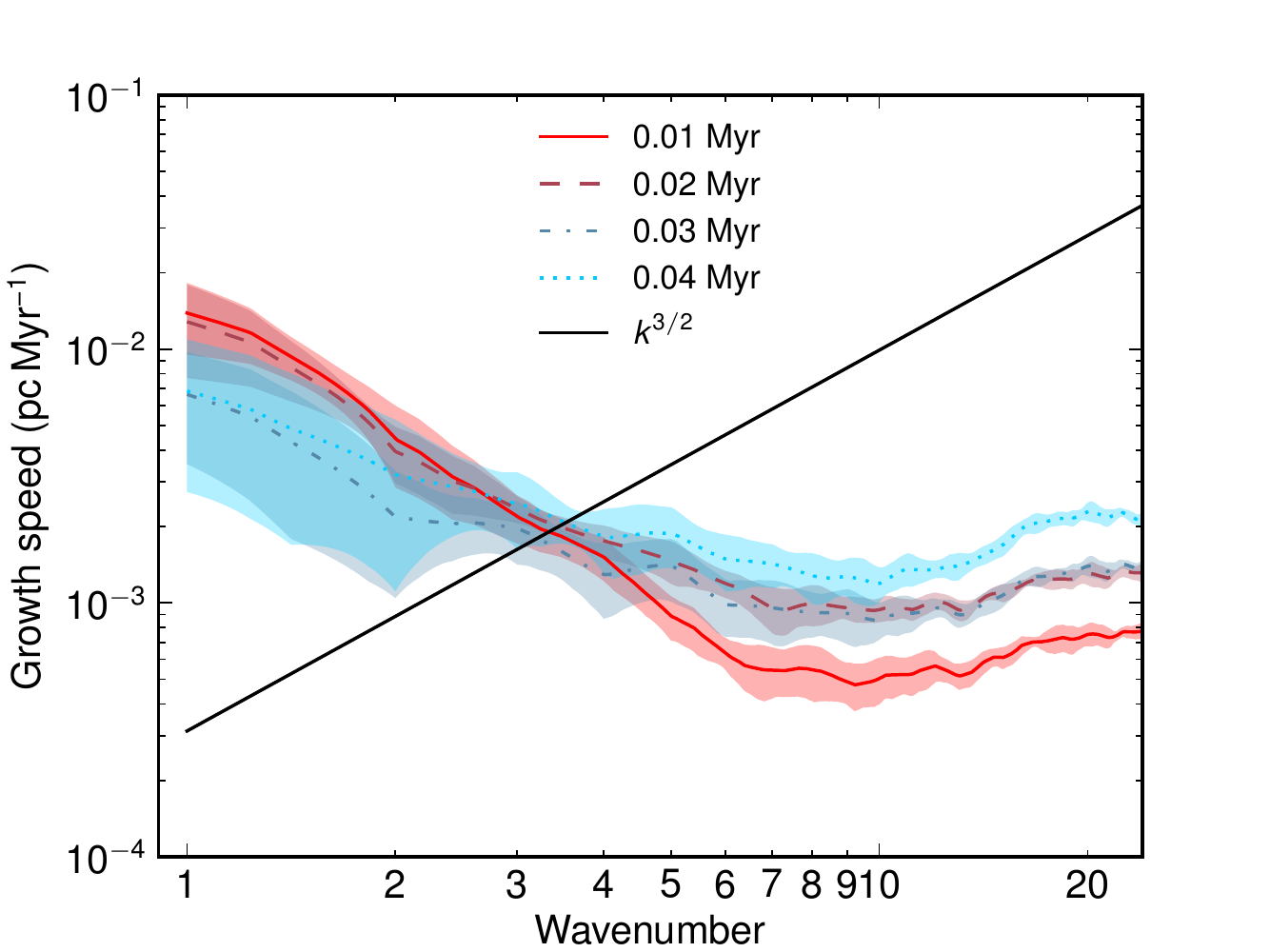}
   \includegraphics[width=\linewidth]{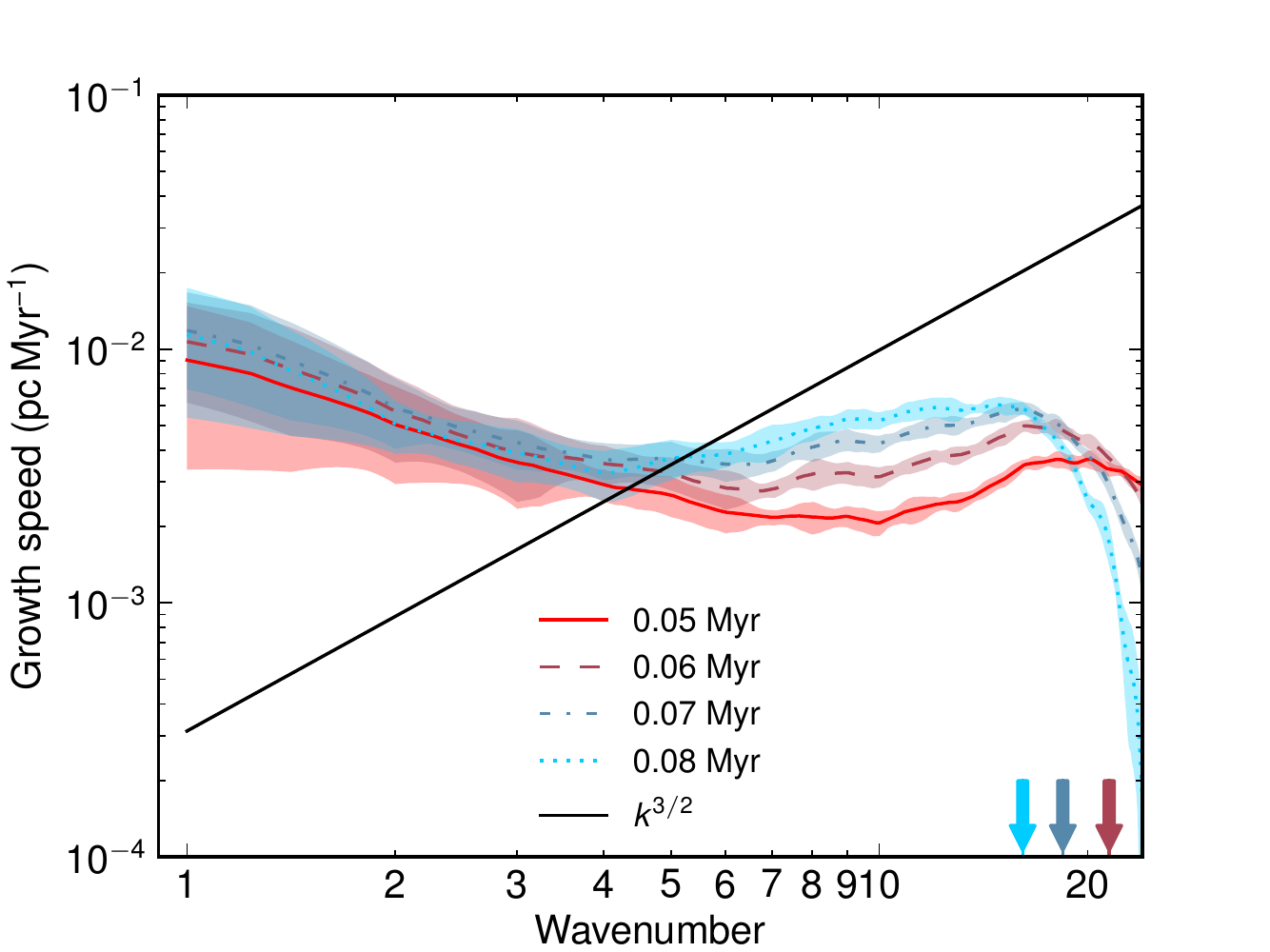}
   \caption[Subsonic turbulence growth speeds]{The growth speeds for simulations with subsonic turbulence as a function of wavenumber, as for Figure~\ref{fig:1D_mono}.}
   \label{fig:2D_lturb}
\end{figure}

\begin{figure}
   \centering
   \includegraphics[width=\linewidth]{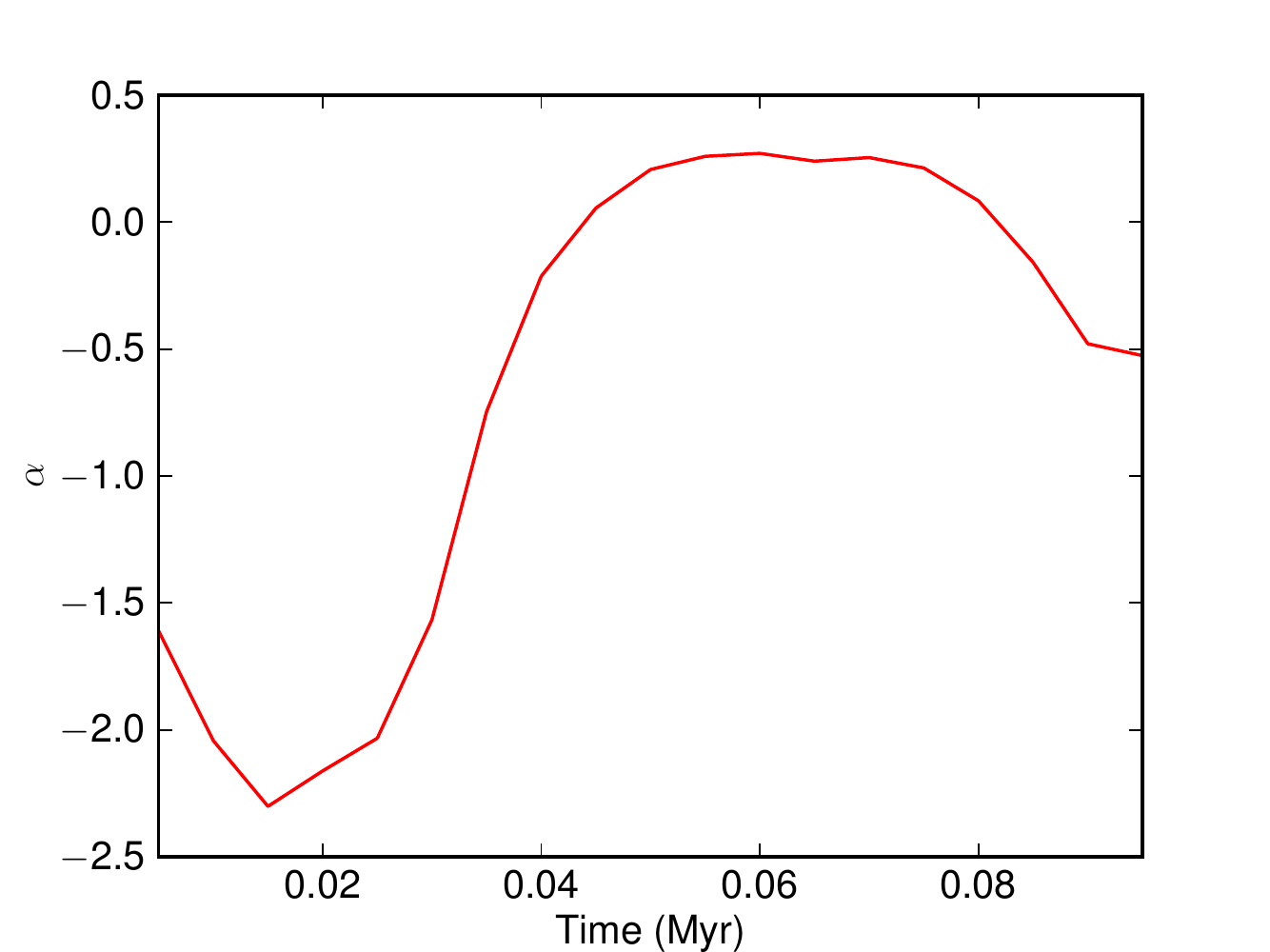}
   \caption[Subsonic turbulence fits]{Indices of power-law fits to growth speeds from simulations with subsonic turbulence, as for Figure~\ref{fig:1D_mono_fits}.}
   \label{fig:2D_lturb_fits}
\end{figure}

Figure~\ref{fig:2D_lturb} shows the growth speeds for simulations with subsonic turbulence. The turbulence remains dominant at smaller wavenumbers; however, the growth speeds are smaller by an order of magnitude. At higher wavenumbers a rise can be seen not too dissimilar in slope to the predictions of \citet{Vishniac1994}. Qualitatively there is a range of wavenumbers over which growth speeds increase with increasing wavenumber until the limit imposed by the thickness of the layer.

Figure~\ref{fig:2D_lturb_fits} shows the index of power-law fits to the growth speeds as a function of wavenumber. Due to the complex nature of these growth speeds, with turbulence dominating at small wavenumbers and the NTSI dominating at intermediate wavenumbers, such fits are difficult. As previously described, for our turbulent simulations we ignore wavenumbers below $k=8$ in our fitting. This excludes the range of wavenumbers still dominated by the turbulence, where growth speeds decrease with increasing wavenumber, and includes only the region where growth speeds are increasing with increasing wavenumber up to the maximum resolvable wavenumber.

As for the supersonic turbulence, the initial index is approximately $-2$. However, this rapidly rises, eventually peaking around 0.7. Although this does not quantitatively match the predicted index of \citet{Vishniac1994}, it is a qualitative match in that higher wavenumbers grow more quickly.

\section{Discussion}

\citet{Hunter_et_al1986} simulated the collision of supersonic flows using a two-dimensional grid code. They found that the dense layer formed became unstable to a `Rayleigh--Taylor-like' instability, and noted that it only occurred in the presence of cooling. This instability, the NTSI, was first analysed for supersonic isothermal collisions by \citet{Vishniac1994}. The NTSI requires a large density contrast between the inflowing and layer gas, and so requires strong cooling to produce an approximately isothermal shock, as found by \citet{Hunter_et_al1986}.

Using a two-dimensional Eulerian code, \citet{BlondinMarks1996} studied the growth of the NTSI in a layer formed between planar flows. These simulations confirmed the analysis of \citet{Vishniac1994} that growth rates were larger at larger wavenumbers. They found the dependence on wavenumber slightly weaker than predicted, finding $\tau^{-1} \propto (k L)^{0.7}$, while we find the dependence slightly stronger than predicted. Qualitatively our results are in good agreement; the NTSI grows exponentially until saturation, which occurs when the layer thickness exceeds half the perturbation wavelength. The layer remains turbulent and bloated, and is much thicker than expected for an idealized isothermal shock, similar to the results of \citet{FoliniWalder2006}.

\citeauthor{BlondinMarks1996} began their simulations with a sinusoidally perturbed slab already in place, and did not use an initial velocity perturbation. However, their simulations, using a similar Mach number of 25, are otherwise similar to our simulations with a one-dimensional monochromatic perturbation. Our results therefore extend their work to two-dimensional perturbations, white noise and turbulent initial conditions, but we find their basic results hold for all cases we examine except those containing turbulence.

 \citet{KleinWoods1998} studied the growth of the NTSI in colliding clouds, while \citet{Hueckstaedt2003} conducted a study of colliding flows similar to that of \citet{BlondinMarks1996}. Both used two-dimensional axisymmetric codes. As in the simulations of \citet{Hunter_et_al1986}, both found that the NTSI requires strong cooling. When using an adiabatic equation of state, \citeauthor{KleinWoods1998} found the NTSI was not excited; they were also able to show that the NTSI did not grow in the absence of initial perturbations, as predicted by \citet{Vishniac1994}. Both sets of simulations are in qualitative agreement with our results; however, neither made a quantitative measurement of the dependence of growth rates on wavenumber.

\subsection{NTSI in the Galaxy}

It is worth considering what astrophysical systems could be susceptible to the NTSI. The NTSI requires a layer with a significant density contrast, bounded by shocks from inflowing gas on both sides. The inflowing gas must therefore be supersonic and must cool rapidly in the shock. We also find that supersonic turbulence can prevent the NTSI from being observed if the growth speeds of turbulent perturbations accreted onto the layer exceed those of the NTSI. Since supersonic turbulence is ubiquitous in the interstellar medium, the NTSI may be replaced in real systems by these turbulent perturbations, which would limit the applicability of our results.

\subsubsection{Colliding clouds}

Colliding molecular clouds are likely to provide the conditions for the NTSI, provided the collisions are not excessively off-centre, and are likely to be the systems to which our simulations are most applicable. The numerical study of colliding clouds has a long history, beginning with the one-dimensional and two-dimensional simulations of \citet{Stone1970a, Stone1970b}. Many further studies followed \citep[e.g.][]{Smith1980, Hausman1981, Gilden1984, Lattanzio_et_al1985}, however many of these would be unable to identify the NTSI due to low resolution, the use of a mirror boundary instead of a second cloud, or the lack of initial perturbations. The exception is the work by \citet{KleinWoods1998} described earlier.

Simulations including cooling showed that such collisions are approximately isothermal \citep[e.g.][]{Stone1970a, Hausman1981, SabanoTosa1985}. Together with the supersonic motions of clouds, this provides the high density contrast required for growth of the NTSI. More recent simulations such as those by \citet{KitsionasWhitworth2007} have tended to focus on the resulting star formation rather than instabilities that occur during the collision.

Both subsonic and supersonic turbulence are common in molecular clouds and cores. Previous studies have not considered the effect of turbulence on the NTSI; our results for subsonic and supersonic turbulence can be applied to the collisions of such turbulent clouds. Further numerical work studying the quantitative growth of the NTSI in colliding clouds is needed.

\subsubsection{Large-scale flows}

Converging flows on galactic scales have been suggested as the precursors to molecular clouds, since such flows can collect material in a relatively dense layer \citep{Heitsch_et_al2005, Heitsch_et_al2006, VazquezSemadeni_et_al2006, VazquezSemadeni_et_al2007a}. Such flows may trigger the NTSI and the Kelvin--Helmholtz instability in the layer, which drives turbulence in the layer.

However, unlike in our simulations the inflowing gas is hot atomic gas. When this gas is compressed it undergoes a thermal instability to form a cold, dense layer. This means that the dense layer and any fragments formed are confined by thermal pressure as well as ram pressure, and so our results may not be directly applicable. Nonetheless the NTSI is an important process if the collision is sufficiently supersonic and the cooling is sufficiently rapid \citep{Heitsch_et_al2005}.

\citet{AuditHennebelle2010} simulated converging flows using either a two-phase medium with a thermal instability or an isothermal equation of state. The velocity of the inflowing gas is the same in both cases, but corresponds to a Mach number of 1.5 for the two-phase medium and of the order of 10 for the isothermal case. The two-phase medium leads to a much more fragmentary structure, with many small objects which are compressed by the surrounding hot medium. The isothermal case is very similar to our simulations of the NTSI, with a more clearly defined layer and many bending modes.

\subsubsection{Colliding winds}

Converging flows occur in the colliding wind binaries examined by \citet{StevensBlondinPollock1992}. They find that the shocked layer produced in such collisions can be isothermal for binaries with a period of less than a few days, particularly for the denser winds of Wolf--Rayet stars. For adiabatic collisions the shocked layer is unstable to the Kelvin--Helmholtz instability, but when the collision is isothermal then a thin shell instability disrupts the layer; \citet{Vishniac1994} identified this as the NTSI. The physical conditions of these collisions is quite different to those of our simulations, but provided the collision is sufficiently radiative the dependence of growth rates on wavenumber should be similar in the region where the collision is close to planar.

\subsubsection{Other converging flows}

The NTSI may also occur in the bow shocks of runaway stars, but our results are less likely to be applicable here; even if the bow shocks are isothermal the dominant instability is likely to be the transverse acceleration instability and not the NTSI \citep*{DganivanBurenNoriega-Crespo1996a, DganivanBurenNoriega-Crespo1996b}. Similarly, the confluence of expanding H\,\textsc{ii} regions \citep{ElmegreenLada1977} or supernovae remnants \citep{Williams_et_al1997} also provide the converging flows required for the NTSI, but may not provide the strong radiative cooling needed for the NTSI.

\subsection{Observations}

We can speculate as to possible observational evidence of the NTSI in real systems. Our results are most closely applicable to molecular cloud collisions and so we focus on this case. We study the growth speeds of perturbations by identifying the position of the dense layer between the clouds. While this may be possible for clouds that collide exactly parallel to the plane of the sky, for most systems it may be easier to observe the line-of-sight velocity of layer perturbations using molecular line observations.

For collisions along our line of sight, the dense layer can be observed using a suitable optically-thin molecular tracer. If the clouds are sufficiently supersonic, their emission should not overlap in velocity space with the dense, mostly subsonic layer. If some of cloud emission is at similar velocities to the layer, then a tracer with a critical density higher than the cloud density but lower than the layer density could be used. This is complicated by the relatively low density contrast between layer and inflowing gas we observe; another possible alternative would be a tracer of shocked gas such as SiO.

Provided suitably high-resolution observations are available then the velocity of the layer provides equivalent information to the layer displacement for our purposes. The velocity across the collision plane can be processed in the same way as we process the layer displacement across the collision plane, except no time derivative is required.

We analyse the growth speeds of perturbations by computing the Fourier spectrum of layer displacements and taking the time derivative of the Fourier spectra; this is equivalent to taking the time derivative of the layer displacements, i.e. the layer velocity, and computing the Fourier spectrum. This will require observations of the velocity across the collision plane; as an example, we estimate the required resolution for a \mbox{10\,pc} cloud at varying distances.

For a minimal Fourier spectrum, we can require a linear resolution of 50 points across the collision plane, which will provide information to a wavenumber of 25. There is likely to be a density gradient across the dense layer from centre to edge due to the spherical nature of the clouds, so we restrict observations to the central \mbox{10\,pc} width and so require a linear resolution of \mbox{0.2\,pc}.

The Atacama Large Millimeter/submillimeter Arrray (ALMA), in its widest configuration, has a maximum spatial resolution of between 6 and \mbox{37\,mas}. Assuming a resolution of \mbox{20\,mas}, this yields our desired linear resolution out to a distance of several megaparsec, and so we easily achieve the required spatial resolution for any target in the Milky Way. More difficult is the required spectral resolution. Since our maximum velocity is of the order of the sound speed, approximately \mbox{$0.2\,\text{km}\,\text{s}^{-1}$}, we require a velocity resolution much smaller than this. The highest spectral resolution available from ALMA is approximately \mbox{$0.01\,\text{km}\,\text{s}^{-1}$}, which only gives a total velocity resolution of about 20 points over the subsonic regime.

We can speculate as to the time that the NTSI could be observed in such a collision. Once the NTSI is saturated, its effects can be seen as a bloated, turbulent layer, but our results will not be applicable to this phase of the layer evolution. In order to observe the slope of the Fourier spectrum, we require smaller wavelength modes to still be growing, which requires that the layer is thinner than half the wavelength. For a wavenumber of 20 across the central \mbox{10\,pc} width, the layer must be thinner than \mbox{0.25\,pc}. Assuming a Mach number of 20, and a density contrast of 30, this will last for approximately \mbox{1\,Myr}. This is similar to the collision time for the clouds, so for this case the NTSI may be growing during the majority of the collision.

\subsection{Magnetic fields}

In this work we do not explore the effect of magnetic fields. Magnetic pressure can support a layer and reduce the density contrast between a layer and the inflowing medium \citep{Stone1970a, SabanoTosa1985}, which should decrease the effectiveness of the NTSI. The magnetized NTSI was studied in detail by \citet{Heitsch_et_al2007} for Mach 4 flows, who found magnetic fields normally weaken the NTSI either by resisting shear, if the field is aligned with the inflow, or by stiffening the equation of state, for fields perpendicular to the inflow. The effect of magnetic fields will depend on the balance of ram pressure and magnetic field strength, and thus may be less important for our large Mach number flows.

\section{Conclusions}

Our simulations have explored the effect of initial perturbations on the growth of the NTSI. We have defined a method to identify the position of a dense shock-compressed layer within a simulation, identify the perturbations within that layer, calculate the growth speeds and so help identify the presence of the NTSI. We have found that, in general, one-dimensional perturbations produce a similar result to two-dimensional perturbations, suggesting that the two-dimensional NTSI is similar to the one-dimensional NTSI analysed by \citet{Vishniac1994}.

We have found that the thickness of the layer increases faster than for the idealized case, but does not depend much on the type of the initial perturbation. This is not a simple problem of resolution, since the layer thickness increases faster in the higher-resolution narrow cross-section simulations than in the much lower-resolution square cross-section simulations. The thickness of the layer does provide a good estimate of the maximum resolvable wavenumber as a function of time, usually to within a factor of two.

We have only considered a single initial collision velocity and single initial density, and therefore have the same range of wavenumbers susceptible to the NTSI in each case. A more complete study might examine the effect of these parameters on the minimum wavenumber susceptible to the NTSI. We have found the original minimum wavenumber limit proposed by \citet{Vishniac1994} to be unreliable, as the layer perturbations can achieve both supersonic speed and high bending angles.

For simulations with one-dimensional monochromatic perturbations, we have qualitatively matched the predictions of \citet{Vishniac1994}, although the slope of the growth speeds is slightly higher than predicted. For two-dimensional monochromatic perturbations, however, the slope of the growth speeds is consistent with \citeauthor{Vishniac1994}'s analytic prediction. It should be noted that this prediction was limited to one-dimensional monochromatic perturbations.

Simulations with one-dimensional white-noise perturbations also match the analytic prediction. Simulations with two-dimensional white-noise perturbations are more complex, but match the general pattern that higher wavenumbers grow more rapidly.

We also consider the effect of turbulence, which is likely to be found in realistic colliding flows. We find that for supersonic turbulence, the turbulence dominates over the NTSI, and the growth speeds reflect the Fourier amplitude spectrum of the turbulence. The growth speeds for the simulations with subsonic turbulence are similar at small wavenumbers, except an order of magnitude weaker. This allows the growth speed of the NTSI to exceed the growth speed of turbulent perturbations over a range of larger wavenumbers, as the turbulence is weaker at larger wavenumbers.

Our results provide a diagnostic to identify the growth of the NTSI in the presence of turbulence. In the absence of the NTSI, growth speeds will decrease with increasing wavenumber if, as is normally assumed, the turbulence is stronger at smaller wavenumbers. If the NTSI is present, there will be a range of wavenumbers over which growth speeds will increase with increasing wavenumber. The limits of this range are the smallest wavenumber where the NTSI grows faster than turbulent perturbations and the largest wavenumber where the NTSI can grow (set by the thickness of the layer).

This diagnostic may be used to identify the presence of the NTSI in colliding flows in more complex and more realistic simulations of astrophysical phenomena, such as colliding molecular clouds. It may also be of use to identify the NTSI in observations of molecular gas in the interstellar medium, where turbulence is believed to be ubiquitous.

\section*{Online materials}

Movies of a representative sample of our simulations can be found online accompanying this paper on the Monthly Notices website. For one-dimensional perturbations, there is a simulation with a $k = 8$ monochromatic perturbation, and a simulation with white noise. For two-dimensional perturbations, there is a simulation with a two-dimensional $k_y = 8$ monochromatic perturbation, and a simulation with two-dimensional white noise perturbations. There is also a simulation with subsonic turbulence and a simulation with supersonic turbulence. There are density cross-sections for all simulations; for simulations with two-dimensional perturbations or turbulence, column density along the $x$-axis is included in separate movie files.

\section*{Acknowledgments}

Plots of SPH data in this work were produced with SPLASH \citep{Price2007}. This research has made use of NASA's Astrophysics Data System. Simulations were performed using the Cardiff University ARCCA (Advanced Research Computing CArdiff) cluster Merlin. Andrew McLeod was funded during this work by an STFC Studentship at Cardiff University. The authors would like to thank the anonymous referee for their helpful and constructive comments.

\footnotesize{
  \bibliographystyle{mn2e}
  \bibliography{NTSI_sim_paper}

\begin{thebibliography}{}

\bibitem[\protect\citeauthoryear{{Audit} \& {Hennebelle}}{{Audit} \&
  {Hennebelle}}{2010}]{AuditHennebelle2010}
{Audit} E.,  {Hennebelle} P.,  2010, \aap, 511, A76

\bibitem[\protect\citeauthoryear{{Blondin} \& {Marks}}{{Blondin} \&
  {Marks}}{1996}]{BlondinMarks1996}
{Blondin} J.~M.,  {Marks} B.~S.,  1996, \na, 1, 235

\bibitem[\protect\citeauthoryear{{Bonnell}, {Dobbs}, {Robitaille} \&
  {Pringle}}{{Bonnell} et~al.}{2006}]{Bonnell_et_al2006}
{Bonnell} I.~A.,  {Dobbs} C.~L.,  {Robitaille} T.~P.,    {Pringle} J.~E.,
  2006, \mnras, 365, 37

\bibitem[\protect\citeauthoryear{{Dgani}, {van Buren} \&
  {Noriega-Crespo}}{{Dgani} et~al.}{1996a}]{DganivanBurenNoriega-Crespo1996a}
{Dgani} R.,  {van Buren} D.,    {Noriega-Crespo} A.,  1996a, \apj, 461, 372

\bibitem[\protect\citeauthoryear{{Dgani}, {van Buren} \&
  {Noriega-Crespo}}{{Dgani} et~al.}{1996b}]{DganivanBurenNoriega-Crespo1996b}
{Dgani} R.,  {van Buren} D.,    {Noriega-Crespo} A.,  1996b, \apj, 461, 927

\bibitem[\protect\citeauthoryear{{Draine} \& {McKee}}{{Draine} \&
  {McKee}}{1993}]{DraineMcKee1993}
{Draine} B.~T.,  {McKee} C.~F.,  1993, \araa, 31, 373

\bibitem[\protect\citeauthoryear{{Elmegreen} \& {Lada}}{{Elmegreen} \&
  {Lada}}{1977}]{ElmegreenLada1977}
{Elmegreen} B.~G.,  {Lada} C.~J.,  1977, \apj, 214, 725

\bibitem[\protect\citeauthoryear{{Folini} \& {Walder}}{{Folini} \&
  {Walder}}{2006}]{FoliniWalder2006}
{Folini} D.,  {Walder} R.,  2006, \aap, 459, 1

\bibitem[\protect\citeauthoryear{{Gilden}}{{Gilden}}{1984}]{Gilden1984}
{Gilden} D.~L.,  1984, \apj, 279, 335

\bibitem[\protect\citeauthoryear{{Gingold} \& {Monaghan}}{{Gingold} \&
  {Monaghan}}{1977}]{GingoldMonaghan1977}
{Gingold} R.~A.,  {Monaghan} J.~J.,  1977, \mnras, 181, 375

\bibitem[\protect\citeauthoryear{{Hausman}}{{Hausman}}{1981}]{Hausman1981}
{Hausman} M.~A.,  1981, \apj, 245, 72

\bibitem[\protect\citeauthoryear{{Heitsch}, {Burkert}, {Hartmann}, {Slyz} \&
  {Devriendt}}{{Heitsch} et~al.}{2005}]{Heitsch_et_al2005}
{Heitsch} F.,  {Burkert} A.,  {Hartmann} L.~W.,  {Slyz} A.~D.,    {Devriendt}
  J.~E.~G.,  2005, \apjl, 633, L113

\bibitem[\protect\citeauthoryear{{Heitsch}, {Slyz}, {Devriendt}, {Hartmann} \&
  {Burkert}}{{Heitsch} et~al.}{2006}]{Heitsch_et_al2006}
{Heitsch} F.,  {Slyz} A.~D.,  {Devriendt} J.~E.~G.,  {Hartmann} L.~W.,
  {Burkert} A.,  2006, \apj, 648, 1052

\bibitem[\protect\citeauthoryear{{Heitsch}, {Slyz}, {Devriendt}, {Hartmann} \&
  {Burkert}}{{Heitsch} et~al.}{2007}]{Heitsch_et_al2007}
{Heitsch} F.,  {Slyz} A.~D.,  {Devriendt} J.~E.~G.,  {Hartmann} L.~W.,
  {Burkert} A.,  2007, \apj, 665, 445

\bibitem[\protect\citeauthoryear{{Heitsch}, {Hartmann} \& {Burkert}}{{Heitsch}
  et~al.}{2008}]{HeitschHartmannBurkert2008}
{Heitsch} F.,  {Hartmann} L.~W.,    {Burkert} A.,  2008, \apj, 683, 786

\bibitem[\protect\citeauthoryear{{Hennebelle}, {Banerjee},
  {V{\'a}zquez-Semadeni}, {Klessen} \& {Audit}}{{Hennebelle}
  et~al.}{2008}]{Hennebelle_et_al2008}
{Hennebelle} P.,  {Banerjee} R.,  {V{\'a}zquez-Semadeni} E.,  {Klessen} R.~S.,
    {Audit} E.,  2008, \aap, 486, L43

\bibitem[\protect\citeauthoryear{{Hubber}, {Batty}, {McLeod} \&
  {Whitworth}}{{Hubber} et~al.}{2011}]{Hubber_et_al2011}
{Hubber} D.~A.,  {Batty} C.~P.,  {McLeod} A.,    {Whitworth} A.~P.,  2011,
  \aap, 529, A27

\bibitem[\protect\citeauthoryear{{Hueckstaedt}}{{Hueckstaedt}}{2003}]{Hueckstaedt2003}
{Hueckstaedt} R.~M.,  2003, \na, 8, 295

\bibitem[\protect\citeauthoryear{{Hunter}, {Sandford}, {Whitaker} \&
  {Klein}}{{Hunter} et~al.}{1986}]{Hunter_et_al1986}
{Hunter} J. J.~H.,  {Sandford} I. M.~T.,  {Whitaker} R.~W.,    {Klein} R.~I.,
  1986, \apj, 305, 309

\bibitem[\protect\citeauthoryear{{Kitsionas} \& {Whitworth}}{{Kitsionas} \&
  {Whitworth}}{2007}]{KitsionasWhitworth2007}
{Kitsionas} S.,  {Whitworth} A.~P.,  2007, \mnras, 378, 507

\bibitem[\protect\citeauthoryear{{Klein} \& {Woods}}{{Klein} \&
  {Woods}}{1998}]{KleinWoods1998}
{Klein} R.~I.,  {Woods} D.~T.,  1998, \apj, 497, 777

\bibitem[\protect\citeauthoryear{{Lattanzio}, {Monaghan}, {Pongracic} \&
  {Schwarz}}{{Lattanzio} et~al.}{1985}]{Lattanzio_et_al1985}
{Lattanzio} J.~C.,  {Monaghan} J.~J.,  {Pongracic} H.,    {Schwarz} M.~P.,
  1985, \mnras, 215, 125

\bibitem[\protect\citeauthoryear{{Lucy}}{{Lucy}}{1977}]{Lucy1977}
{Lucy} L.~B.,  1977, \aj, 82, 1013

\bibitem[\protect\citeauthoryear{{Mac Low}, {Klessen}, {Burkert} \&
  {Smith}}{{Mac Low} et~al.}{1998}]{MacLow_et_al1998}
{Mac Low} M.,  {Klessen} R.~S.,  {Burkert} A.,    {Smith} M.~D.,  1998,
  Physical Review Letters, 80, 2754

\bibitem[\protect\citeauthoryear{{Mac Low} \& {Klessen}}{{Mac Low} \&
  {Klessen}}{2004}]{MacLowKlessen2004}
{Mac Low} M.-M.,  {Klessen} R.~S.,  2004, Reviews of Modern Physics, 76, 125

\bibitem[\protect\citeauthoryear{{Price}}{{Price}}{2007}]{Price2007}
{Price} D.~J.,  2007, \pasa, 24, 159

\bibitem[\protect\citeauthoryear{{Price} \& {Monaghan}}{{Price} \&
  {Monaghan}}{2004}]{PriceMonaghan2004}
{Price} D.~J.,  {Monaghan} J.~J.,  2004, \mnras, 348, 139

\bibitem[\protect\citeauthoryear{{Sabano} \& {Tosa}}{{Sabano} \&
  {Tosa}}{1985}]{SabanoTosa1985}
{Sabano} Y.,  {Tosa} M.,  1985, \apss, 115, 85

\bibitem[\protect\citeauthoryear{{Smith}}{{Smith}}{1980}]{Smith1980}
{Smith} J.,  1980, \apj, 238, 842

\bibitem[\protect\citeauthoryear{{Springel} \& {Hernquist}}{{Springel} \&
  {Hernquist}}{2002}]{SpringelHernquist2002}
{Springel} V.,  {Hernquist} L.,  2002, \mnras, 333, 649

\bibitem[\protect\citeauthoryear{{Stevens}, {Blondin} \& {Pollock}}{{Stevens}
  et~al.}{1992}]{StevensBlondinPollock1992}
{Stevens} I.~R.,  {Blondin} J.~M.,    {Pollock} A.~M.~T.,  1992, \apj, 386, 265

\bibitem[\protect\citeauthoryear{{Stone}}{{Stone}}{1970a}]{Stone1970a}
{Stone} M.~E.,  1970a, \apj, 159, 277

\bibitem[\protect\citeauthoryear{{Stone}}{{Stone}}{1970b}]{Stone1970b}
{Stone} M.~E.,  1970b, \apj, 159, 293

\bibitem[\protect\citeauthoryear{{V{\'a}zquez-Semadeni}, {Ryu}, {Passot},
  {Gonz{\'a}lez} \& {Gazol}}{{V{\'a}zquez-Semadeni}
  et~al.}{2006}]{VazquezSemadeni_et_al2006}
{V{\'a}zquez-Semadeni} E.,  {Ryu} D.,  {Passot} T.,  {Gonz{\'a}lez} R.~F.,
  {Gazol} A.,  2006, \apj, 643, 245

\bibitem[\protect\citeauthoryear{{V{\'a}zquez-Semadeni}, {G{\'o}mez},
  {Jappsen}, {Ballesteros-Paredes}, {Gonz{\'a}lez} \&
  {Klessen}}{{V{\'a}zquez-Semadeni} et~al.}{2007}]{VazquezSemadeni_et_al2007a}
{V{\'a}zquez-Semadeni} E.,  {G{\'o}mez} G.~C.,  {Jappsen} A.~K.,
  {Ballesteros-Paredes} J.,  {Gonz{\'a}lez} R.~F.,    {Klessen} R.~S.,  2007,
  \apj, 657, 870

\bibitem[\protect\citeauthoryear{{Vishniac}}{{Vishniac}}{1994}]{Vishniac1994}
{Vishniac} E.~T.,  1994, \apj, 428, 186

\bibitem[\protect\citeauthoryear{{Williams}, {Chu}, {Dickel}, {Beyer}, {Petre},
  {Smith} \& {Milne}}{{Williams} et~al.}{1997}]{Williams_et_al1997}
{Williams} R.~M.,  {Chu} Y.-H.,  {Dickel} J.~R.,  {Beyer} R.,  {Petre} R.,
  {Smith} R.~C.,    {Milne} D.~K.,  1997, \apj, 480, 618

\end{thebibliography}
}

\bsp

\label{lastpage}

\end{document}